\documentstyle[aps,pre,epic,eepic,graphicx,floats,amsfonts]{revtex}

\begin{document}
                                
\twocolumn

\title{Dimension of interaction dynamics}
\author{Daniel W\'{o}jcik${}^{\dagger*}$, Andrzej
  Nowak${}^{\ddagger*}$, Marek Ku\'{s}${}^{\dagger*\S}$}   
\address{${}^{\dagger}$Center for Theoretical Physics, Polish Academy
  of Sciences, Al. Lotnik\'{o}w 32/46, 02-668 Warszawa, Poland} 
\address{${}^{\ddagger}$Institute for Social Studies, ul. Stawki 5/7,
  00-183 Warszawa, Poland} 
\address{${}^*$College of Science, Al. Lotnik\'{o}w 32/46, 02-668
  Warszawa, Poland}  
\address{${}^\S$ Laboratoire Kastler-Brossel, Universit\'e Pierre et
  Marie Curie, 4. pl. Jussieu, 75252 Paris, France} 
\date{\today}

\maketitle

\begin{abstract}
  A method allowing to distinguish interacting from
  non-interacting systems based on available time series is proposed
  and investigated. Some facts concerning generalized Renyi dimensions
  that form the basis of our method are proved.
  We show that one can find the dimension of the part of the attractor
  of the system connected with interaction between its parts. We use
  our method to distinguish interacting from non-interacting systems
  on the examples of logistic and H\'enon maps. A~classification of all
  possible interaction schemes is given. 
\end{abstract}

\draft


\section{Introduction}

Given two time series can one tell if they originated from interacting
or non-interacting systems? We show that with the help of embedding
methods\cite{Packard80}, Takens theorem\cite{Takens80,Sauer91a} and
some facts concerning Renyi dimensions which we prove, one can succeed
in case of chaotic systems. Moreover, one can quantify the common part
of the dynamics, which we call dynamics of interaction. 

It happens sometimes, especially in simple systems like electronic
circuits or coupled mechanical oscillators, that one knows whether the
systems under investigation are coupled or not, what is the direction
and sometimes strength of the coupling. However, there are many complex
phenomena in nature where one is unable to verify directly the existence of
coupling between parts of the system in which the phenomenon takes place.  

Especially in complex spatiotemporal systems, like fluid systems,
brain, neuronal tissue, social systems etc. one often faces the
problem of characterization of interdependence of parts of the system
of interest and quantifying the strength of interactions between the
parts. 

Recent research in neurology, for example, has shown that temporal 
coordination between different, often distant neural assemblies plays a 
critical role in the neurophysiological underpinnings of such cognitive 
phenomena as the integration of features in object representation (cf. 
\cite{suave99n} for a review) and the conscious experience of stimuli
\cite{tononi98n}.  The critical empirical question, therefore, is
which of the neural assemblies synchronize their activity.  Since
coordination may take many forms, including complex non-linear
relations, simple correlational methods may not be sufficient to detect
it.  The detection of  
nonlinear forms of coordination is also critically important for issues in 
cognitive science \cite{port95n}, developmental psychology 
\cite{thelen94n}, and social psychology \cite{nowak98n,vallacher94n}.

The method traditionally used for this purpose is correlation
analysis. Given two time series one studies their autocorrelation
functions and cross-correlations. Large cross-correlations are
usually attributed to large interdependence between the parts. Small
cross-correlations are considered as the signature of independence of
the variables. 

Unfortunately, the linear time series analysis gives meaningful
results only in case of linear systems or sto\-chas\-tic time series. It
is well-known that spectral analysis alone cannot discriminate between
low-dimensional  non-linear deterministic systems and
stochastic sys\-tems\cite{Kantz97c}, even though the properties of the two kinds of systems
are different. 

Methods based on entropy measures represent one viable approach for 
detecting nonlinear relations between the activity of different neural 
assemblies  \cite{tononi98n}.

Recently another approach based on nonlinear mutual prediction has been
proposed and used in an ex\-pe\-ri\-ment.
Pecora, Caroll and Heagy\cite{Pecora95} developed a statistics to
stu\-dy the topological nature of functional relationship between
coupled systems. Schiff et al.\cite{Schiff96} used it as a~basis of
their method. The idea is as follows: if there exists a functional
relationship between 
two systems, it is possible to predict state of one system from the
known states of the other. This happens if the coupling between two
systems is strong enough so that generalized synchronization
occurs\cite{Pecora95,Rulkov95,Pecora97}. The average normalized mutual
prediction error is used to quantify the strength and directionality
of the coupling\cite{Schiff96}.

The method we introduce in the present paper does not assume
generalized synchrony. We introduce the notion of the {\em dimension
of interaction}, which measures the size of the dynamics responsible
for the coupling between the two systems. More precisely, it is the
dimension of the part of the attractor of the whole system, which is
acted on by the dynamics of both subsystems. We also show how to
obtain information concerning the strength and directionality of the
coupling.

The idea is, in fact, very simple. Given two time series from
subsystems of interest we construct another one which probes the whole
system, for instance adding the two series. If the subsystems do not
interact, dimension of the whole system is the sum of the dimensions
of the two subsystems, all of which can be estimated from data. On the
other hand, if the subsystems have some common degrees of freedom,
dimension of the whole system will be smaller than the sum of the
dimensions of the two subsystems.

Our method can also be used to find out if two response systems have
a common driver. We discuss this application in
Section~\ref{sec:method}. 

The structure of the paper is as follows. In Section%
~\ref{sec:theory} we recall the definition of the Renyi dimensions and
formulate three theorems which form the basis of our method. The,
rather straightforward, proofs have been relegated to
Appendix~A, since they are not crucial for understanding the
method itself and can be omitted by readers whose main interest is in
applications. 
We formulate our method in section~\ref{sec:method}.
Classification of all the possible interaction schemes is given in
Section~\ref{sec:class}. A simple way of verifying the kind and
direction of the coupling is provided. 
Results from the simulations of coupled logistic and  H\'enon maps 
are collected in Section~\ref{sec:results}. 
Final comments and outlook are given in the last section.

\section{Theoretical considerations}
\label{sec:theory}

Our method presented in Section~\ref{sec:method} is based on three
theorems relating dimensions of subsystems to the dimension of the
whole system. The first one states the intuitively obvious fact that
the dimension of a system consisting of two non-interacting parts is
the sum of the dimensions of the subsystems. A less trivial Theorems
2 and 3 establish interdependencies among the dimensions of the system
and its interacting parts. 
Before we state our theorems we shall recall the definition of the
Renyi dimensions.    

\subsection{Renyi dimensions}
\label{sec:renyi}

It is at present generally accepted that a lot of objects, both in the
real physical space and in the phase space, are multifractals
\cite{Mandelbrot82,Meakin98da,Pesin97dc,Olsen95da}. This means they
can be described by (statistically) self-similar probability measures. This
usually implies that they can be decomposed into a (infinite) number of objects of 
different Haussdorff dimensions, or, equivalently, they have
non-trivial multifractal spectra of dimensions. 

The Renyi dimensions\cite{Rnyi71} have drawn attention of physicists
and mathematicians after publication of the papers by Grassberger,
Hentschel and Procaccia
\cite{Grassberger83,Hentschel83,Grassberger83c,Grassberger83d}. 
For a probability measure $\mu$ on a $d$-dimensional space $U$ one
takes a partition of $U$ into small cells of equal linear size
$\varepsilon$ (equal volume $\varepsilon^d$). One 
defines the Renyi dimensions\footnote{An equivalent description of
  multifractal measures is $f_\alpha$  
  spectrum\cite{Badii85,Halsey86,Frisch85}. A thorough discussion of the
  properties of $D_q$ and $f_\alpha$ spectra falls beyond the scope of
  this paper. Some good reviews of these with the discussion of the
  abundant literature on multifractals can be found e.g. in 
  \cite{Meakin98da,Pesin97dc,Olsen95da,Paladin87,Tl88,Evertsz92da,Ott93,Beck97,Falconer97da}.  
  Mathematically precise definitions of multifractal spectra can be
  found in \cite{Pesin97dc,Olsen95da}.}
 as 
\begin{equation}
  \label{renyi}
  D_q(\mu)  := \left\{
    \begin{array}[c]{ll}
      \lim_{\varepsilon\rightarrow0} \frac{1}{q-1} 
      \frac{\log \sum_i p_i^q}{\log \varepsilon}, &\quad {\rm for}\, q\in
      {\mathbb{R}}\setminus\{1 \}\\
      \lim_{\varepsilon\rightarrow0} \frac{1}{q-1} 
      \frac{\sum_i p_i \log p_i}{\log \varepsilon}, &\quad {\rm for}\, q=1,
    \end{array}
  \right.
\end{equation}
where
\[
p_i=\mu(i-\mbox{\rm th cell})=\int_{i-\mbox{\rm th cell}} d\mu(x),
\]
and the sum is taken over all cells with $p_i\neq 0$.

Of particular importance are $D_0$ --- the box-counting dimension, usually
equal to the Hausdorff dimension\cite{Mandelbrot82,Hausdorff19,Farmer83}, $D_1$ ---
the information dimension or the dimension of the
measure\cite{Rnyi71,Farmer82c,Badii84,Badii85,Grassberger85}, which
describes how the entropy $-\sum_i p_i \log p_i$ increases with the
change of the scale, and $D_2$ --- the correlation
dimension\cite{Grassberger83c,Grassberger83d,Kantz93c}, which can be
most easily extracted from data, usually treated as a lower estimate
of $D_1$ since $D_{q_1}\leq D_{q_2}$ for $q_1>q_2$.

Generalized dimensions are defined for all real $q$, however in proofs
we shall restrict our attention to the case $q\geq 1$. We are
particularly interested in $q=1$ and $q=2$. 
  
\subsection{Non-interacting systems}
\label{sec:nonsyst}

Consider two non-interacting dynamical systems
$(U_1,\varphi_1,\mu_1)$, $(U_2,\varphi_2,\mu_2)$, where $U_i\subset
{\mathbb{R}}^{n_i}$ is the phase space, $\varphi_i$ is a flow or a
map on $U_i$, and $\mu_i$ is an ergodic $\varphi_i$-invariant natural
measure on $U_i$. 

Below we shall concentrate on the case of continuous systems. Changes
needed for the discrete time case are mostly notational.

By natural measure $\mu$ we mean 
\begin{equation}
\label{measure}
\mu=\mu(x_0):=\lim_{T\rightarrow\infty} \frac1T \int_0^T
\delta(x-\varphi_t(x_0))dt; 
\end{equation}
in the weak sense for $\mu$-almost every $x_0$ (one typically thinks
of some physical measure, like Sinai-Ruelle-Bowen
measure\cite{Eckmann85}); $\mu_i(U_i)=1$. The limit in the weak sense
means that if we integrate $\mu(x_0)$ with a continuous function $f$ on
$U$ the limit (\ref{measure}) exists and is $\mu$-almost everywhere
independent of $x_0$, or in other words, the average of $f$ along a
typical trajectory is independent of the trajectory, thus time
averages are equal to ensemble averages. We are interested in such
measures that the set of $x_0$ for which $\mu(x_0)=\mu$ has a non-zero
Lebesgue measure.    

The composite non-interacting system has a product structure
$(U_1\times U_2, \varphi_1\times\varphi_2, \mu_1\times\mu_2)$. Its
dynamics can be written as  
\[
\left\{
  \begin{array}[c]{rcl}
    {\mathbf u}_1(t) & = & \varphi_1({\mathbf u}_1(0),t),\\
    {\mathbf u}_2(t) & = & \varphi_2({\mathbf u}_2(0),t).
  \end{array}
\right.
\]

{\bf Theorem 1}
{\it Suppose $D_q(\mu_1),D_q(\mu_1),D_q(\mu_1\times \mu_2)$ exist. Then 
\begin{equation}
  \label{cart}
  D_q(\mu_1\times\mu_2)=D_q(\mu_1)+D_q(\mu_2).
\end{equation}
}

This means, as should be intuitively obvious, the dimensions of
non-interacting subsystems add up to the dimension of the whole
system. The proof is given in Appendix~A. It follows from extensivity
of Renyi entropies.

This is in fact one of the long-standing problems in the dimension
theory, namely finding the conditions under which the equality holds
for various dimensions for arbitrary measures. Some results for
Olsen's version of multifractal formalism with a discussion of
previous results can be found in \cite{Olsen96da}.  
 
\subsection{Interacting systems}
\label{sec:intersyst}

Take two interacting subsystems $U_1$ and $U_2$ of system $U$. 
It may happen that all the variables in $U_1$ couple with all those in
$U_2$ but this is not necessary.
For many-dimensional systems the structure of the equations of
dynamics can be very complicated. 

Consider the following decomposition of variables of $U_i$. Let
${\mathbf y}_1$ be the largest set of variables in $U_1$ satisfying
the condition that if you change their state whatsoever, it will not
influence the future evolution of $U_2$. Similarly define ${\mathbf
  y}_2$. Put all the other variables of $U_1,U_2$ in vector ${\mathbf
  x}$. They form a dynamical system $V$ --- the part of the whole
system which is responsible for the interaction.
Then the dynamics of the whole system $U$ can be written as 
\begin{equation}
\label{eq:doubledrive} 
\left\{
  \begin{array}[c]{rcl}
    \dot{{\mathbf x}} & = & f({\mathbf x}),\\
    \dot{{\mathbf y}}_1 & = & g_1({\mathbf x},{\mathbf y}_1),\\
    \dot{{\mathbf y}}_2 & = & g_2({\mathbf x},{\mathbf y}_2).
  \end{array}
\right.
\end{equation}
Thus the dynamics of the interacting systems $U_1$ and $U_2$ is
formally equivalent to dynamics of three systems:
$X$ (interaction part) driving $Y_1$ and $Y_2$. We pursue this analogy
deeper in the next section. An example when such decomposition
arises naturally is given in Appendix~B. 

Let $\mu_U, \mu_1, \mu_2, \mu_V$ be natural measures of dynamical
systems, respectively, $U, U_1, U_2, V$.

{\bf Theorem 2} 
{\em Suppose $D_1(\mu_1)$, $D_1(\mu_2)$, $D_1(\mu_V)$,
  $D_1(\mu_U)$ exist. Then 
  \begin{equation}
    \label{eq:inter}
    D_1(\mu_V) \leq d_{\sf int}:= D_1(\mu_1) + D_1(\mu_2) - D_1(\mu_U).
  \end{equation}
(We shall call $d_{\sf int}$ {\em dimension of interaction}). The
equality holds when ${\mathbf y}_1$ and ${\mathbf y}_2$ are
asymptotically independent. 
}

Asymptotical independence means essentially lack of
generalized synchronization between the ${\mathbf y}$s and their
common driver ${\mathbf x}$. We relegate further discussion to
Appendix~A, where we make this condition precise and show where it is
needed\footnote{One would like to establish a similar inequality in case of other Renyi
dimensions, however, in general 
\[
D_q(\mu_V) + D_q(\mu_U) - D_q(\mu_1) - D_q(\mu_2)
\]
can have arbitrary sign (cf. Appendix~A). Nevertheless, we expect this
difference for typical physical systems to be small in comparison with
the dimensions involved.}.

If we think of dimensions as estimates on the number of degrees of
freedom, the Theorem~2 means intuitively that if the system
can be considered as composed of interacting parts, some of the degrees of
freedom --- perhaps even all --- are common for both of the parts. Therefore,
dimension of the whole system is equal to the sum of the number of the common
degrees of freedom, those degrees of freedom which belong to 
$U_1$ and do not belong to $U_2$, and the other way round. Thus if we
add the dimensions of the subsystems $U_1$ and $U_2$, we count the
common degrees of freedom twice. We must therefore subtract them if we
want to get dimension of the whole system $U$.

In the above theorem we show that this intuition can be made
precise {\em only} in case of the information dimension $D_1$ and with
an additional assumption.
The notion of the dimension of interaction we define in equation~(\ref{eq:inter})
is crucial for our method.

In the special case, when one (or both) of $U_i=V$, 
(all the variables of $U_i$ couple with some of the variables of the
other subsystem), say $U_2=V$, 
we may establish  

{\bf Theorem 3}
{\em  Suppose $D_q(\mu_1)$, $D_q(\mu_2)$, $D_q(\mu_V)$
  exist and $k_2=n_2$. Then 
  \begin{equation}
    \label{inter2}
    D_q(\mu_V) = d_q^{\sf int}:= D_q(\mu_1) + D_q(\mu_2) -
    D_q(\mu_U). 
  \end{equation}
}

The proof is obvious, for in this case $U_2\equiv V$ and $U_1=U$.
This also means that the above intuitions in this case are precise for
arbitrary generalized dimensions and no further assumptions are needed.   

The generalized dimensions of interaction $ d_q^{\sf int}$ are
estimates on the number of effective degrees of freedom responsible
for the interaction between the parts of the system under study. Of
most interest are $  d_1^{\sf int}\equiv d_{\sf int}$, which has the
best analytical properties, and $d_2^{\sf int}$, which can be most
reliably estimated from data.

Note that 
\begin{eqnarray}
  \max \{d_q^{\sf int},D_q(\mu_V)\} & \leq & \min 
  \{ D_q(\mu_1),D_q(\mu_2)\} \nonumber \\   
  & \leq & \max \{ D_q(\mu_1),D_q(\mu_2)\} \label{diminequal}\\
  & \leq & D_q(\mu_U).\nonumber
\end{eqnarray}
Furthermore, for $q=1$ one can show
\[
0 \leq D_1(\mu_V) \leq d_1^{\sf int}.
\]
We conjecture $d_q^{\sf int}\geq 0$ also for  $q>1$.

\section{The method}
\label{sec:method}

Suppose we are given two time series measured in subsystems $U_1$ and
$U_2$ of system $U$ whose
structure and interdependence we do not know, e.g. signals gathered on
two electrodes placed in not too far away portions of brain, or
measurements of velocity or temperature in various parts of moderately
turbulent fluid. We would like to know, if the e\-qua\-tions governing the
dynamics of both of these variables are coupled or not, how many
degrees of freedom are common and what is the direction of the coupling.
   
Let $X_i$ be a function on $U_i$, i.e. $X_i :
U_i\rightarrow\mathbb{R}$. The time series we measure are
$x_1(n):=X_1({\mathbf u}_1(t_n))$ and $x_2(n):=X_2({\mathbf u}_2(t_n))$.  
Let $Y:\mathbb{R}^{\rm 2}\rightarrow\mathbb{R}$ be a smooth function
nontrivially depending on both variables\footnote{\label{foot} For
  finite noisy time series some functions are 
  better than other. In practice we used five different functions
  $Y(x,y)$, namely $x+y$, $x\cdot y$, $\sin(x)\cos(y)$, $x\exp(y)$,
  $2x-y$, to calculate dimension of the system $D_q(\mu_U)$, and
  averaged the results. The variance of the obtained five estimates
  was usually small.

  The above functions were not chosen for their particularly good
  numerical properties but rather to verify that the results obtained
  depend only weekly on the choice of the function $Y$.}. We construct
another time series $y(n)=Y(x_1(n),x_2(n))$. Thus $Y(X_1,X_2)$ is a
function on $U$. 

Using time delay method\cite{Packard80,Takens80} we can reconstruct
the dynamics of the systems $U_i$ and $U$ from $x_i(n)$ and
$y(n)$. Namely, for a given {\em delay\/} $\tau$ and {\em embedding
  dimension\/} $N$ we construct {\em delay vectors}
\[
{\tilde{\mathbf u}}_1(n)=(x_1(n),x_1(n-\tau),\dots,x_1(n-(N-1)\tau));
\]  
the construction of ${\tilde{\mathbf u}}_2$ from $x_2$ and
${\tilde{\mathbf u}}$ from $y$ is similar. 

If $N>2 D_0(\mu_1)$, for all reasonable delays, for infinite
not-too-sparsely probed time series, the Takens
theorem\cite{Takens80,Sauer91a} guarantees ${\tilde{\mathbf u}}_1(n)$
is an embedding of the original invariant set in $U_1$. To calculate
dimensions it is even enough to take $N> D_0(\mu_1)$
\cite{Sauer93a,Sauer97b}. It is generally believed that also for
finite but not too short and not too noisy time series the above
construction gives occasionally a reasonable estimate on the original
dynamics. For a detailed discussion of these issues the reader should
consult the relevant literature,
e.g. \cite{Kantz97c,Abarbanel93a,Abarbanel96,Schreiber99,Casdagli91a}.
We disregard the practical problems until section~\ref{sec:results}
where we show some numerical results. For the time being we discuss
clean infinite time series.

Having reconstructed the attractors we can estimate their generalized
dimensions and calculate the {\em generalized dimensions of interaction}  
\begin{equation}
  \label{eq:dint}
  d_q^{\sf int}:=D_q(\mu_1) + D_q(\mu_2) - D_q(\mu_U).
\end{equation} 

It is also convenient to consider normalized dimensions of interaction:
\begin{equation}
  \label{eq:normdim}
  \begin{array}{rcl}
    m_1^q & := &  d_q^{\sf int}/D_q(\mu_1),\\
    m_2^q & := &  d_q^{\sf int}/D_q(\mu_2),\\
    m_U^q & := &  d_q^{\sf int}/D_q(\mu_U).
  \end{array}
\end{equation}
From the values of $m_i^q$ we can infer the information we need. All the
possible cases are described in the next section.
Note that if $m_i^q\neq 0$, they  satisfy
\[
\frac{1}{m_1^q} + \frac{1}{m_2^q} - \frac{1}{m_U^q}=1.
\]
From~(\ref{diminequal}) we also have
\[
0 \leq m_U^q \leq m_1^q,m_2^q \leq 1,
\]
which provides us with a tool to check consistency of results.

Before we present the classification of all the possible schemes of
interaction let us discuss heuristically four simple examples.

\begin{enumerate}
\item[I] If $U_1,U_2$ are uncoupled, the variables we see through $x_1,x_2$ are
  different, thus $\mu_U=\mu_1\times\mu_2$. Therefore, from Theorem 1,
  $d_q^{\sf int}=0$, as it should be for any reasonable definition of
  dimension of interaction for non-interacting systems. 
  
\item[II] Consider now a system $U$ consisting of three isolated systems $V_i$,
  which we cannot observe separately, however, but rather through $U_1$ and
  $U_2$, e.g. measuring $X_1({\mathbf v}_1,{\mathbf v}_2)$ and
  $X_2({\mathbf v}_2,{\mathbf v}_3)$.
  \begin{figure}[htbp]
    \begin{center}
      \leavevmode 
      \setlength{\unitlength}{0.00033333in}
      \begingroup\makeatletter\ifx\SetFigFont\undefined%
      \gdef\SetFigFont#1#2#3#4#5{%
        \reset@font\fontsize{#1}{#2pt}%
        \fontfamily{#3}\fontseries{#4}\fontshape{#5}%
        \selectfont}%
      \fi\endgroup%
      {\renewcommand{\dashlinestretch}{30}
        \begin{picture}(7216,3029)(0,-10)
          \put(3608,1507){\ellipse{1800}{1800}}
          \put(1208,1507){\ellipse{1800}{1800}}
          \put(6008,1507){\ellipse{1800}{1800}}
          \put(2408,1507){\ellipse{4800}{3000}}
          \put(4808,1507){\ellipse{4800}{3000}}
          \put(908,1507){\makebox(0,0)[lb]{\smash{{{\SetFigFont{10}{24.0}{\rmdefault}{\mddefault}{\updefault}$V_1$}}}}}
          \put(3308,1507){\makebox(0,0)[lb]{\smash{{{\SetFigFont{10}{24.0}{\rmdefault}{\mddefault}{\updefault}$V_2$}}}}}
          \put(5708,1507){\makebox(0,0)[lb]{\smash{{{\SetFigFont{10}{24.0}{\rmdefault}{\mddefault}{\updefault}$V_3$}}}}}
          \put(1808,307){\makebox(0,0)[lb]{\smash{{{\SetFigFont{10}{24.0}{\rmdefault}{\mddefault}{\updefault}$U_1$}}}}}
          \put(4508,307){\makebox(0,0)[lb]{\smash{{{\SetFigFont{10}{24.0}{\rmdefault}{\mddefault}{\updefault}$U_2$}}}}}
        \end{picture}
        }
      \caption{Simple interaction}
      \label{fig:simple}
    \end{center}
  \end{figure}
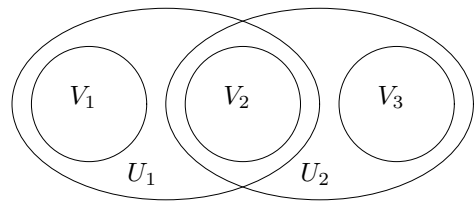
  Reconstructing dynamics from time series of $X_1$ and $X_2$ we
  expect to obtain
  \begin{eqnarray*}
    D_q(\mu_1) & = & D_q(\mu_{V_1})+D_q(\mu_{V_2}), \\    
    D_q(\mu_2) & = & D_q(\mu_{V_2})+D_q(\mu_{V_3}).      
  \end{eqnarray*}
  With a typical function $Y(x_1,x_2)$ we obtain time series $y(n)$
  from which we estimate
  \[
  D_q(\mu_U)= D_q(\mu_{V_1})+D_q(\mu_{V_2})+D_q(\mu_{V_3}).      
  \]
  Since dynamics of $V_2$ is responsible for the interaction between $U_1$
  and $U_2$, we want to call the {\em dimension of
    interaction} dimension of $\mu_{V_2}$. According to the definition
  (\ref{eq:dint}) we have 
  \begin{eqnarray*}
    d_q^{\sf int} & = & D_q(\mu_1)+D_q(\mu_2)-D_q(\mu_U)\\
    & = & D_q(\mu_{V_1})+D_q(\mu_{V_2})+D_q(\mu_{V_2})+D_q(\mu_{V_3})+\\
    & & -[ D_q(\mu_{V_1})+D_q(\mu_{V_2})+D_q(\mu_{V_3})]\\
    & = & D_q(\mu_{V_2}).
  \end{eqnarray*}

\item[III] Consider now the general situation described in
  section~\ref{sec:intersyst}.  
  Reconstructing dynamics from time series of typical variables from
  systems $U_1$ and $U_2$, say $x_1(n)$ and $x_2(n)$
  we get
  \begin{eqnarray*}
    D_q(\mu_1) & \geq &  D_q(\mu_V), \\
    D_q(\mu_2) & \geq &  D_q(\mu_V).
  \end{eqnarray*}
  For a typical function $Y(x_1,x_2)$ we obtain 
  \begin{eqnarray*}
    \lefteqn{\max\{D_q(\mu_1),D_q(\mu_2)\}}\\
    & \leq & D_q(\mu_U)\\
    &\leq&  D_q(\mu_V)+ (D_q(\mu_1)- D_q(\mu_V)) +\\
    && (D_q(\mu_2)- D_q(\mu_V)),
  \end{eqnarray*}
  where $D_q(\mu_1)- D_q(\mu_V)$ quantifies number od degrees of
  freedom in $U_1$ not coupled to $U_2$. From this we conclude
  \begin{eqnarray*}
    0 < D_q(\mu_V)&\leq& d_q^{\sf int}\\
    & = &D_q(\mu_1)+D_q(\mu_2)-D_q(\mu_U)\\
    & \leq & \min\{D_q(\mu_1),D_q(\mu_2)\},
  \end{eqnarray*}
  the difference between $D_q(\mu_V)$ and $d_q^{\sf int}$ depending on
  the strength of synchronization between $U_1$ and $U_2$. 

\item[IV] As the last example we shall take a system $X$ driving two
  response systems $Y_1$ and $Y_2$. Suppose we also have a second copy
  of this setup, namely drive $X'$ with response systems $Y_1'$ and
  $Y_2'$. We collect simultaneously four time series of some variable
  from all of the response systems. Now we choose randomly two of them
  and want to know if the systems they come from had a common driver. 

  It is easy to check that if they had, then $d_q^{\sf int}$ is
  approximately the dimension of the driver $D_q(\mu_X)>0$. If they had
  different drivers, then $d_q^{\sf int}=0$.  

\end{enumerate}

Summarizing, from measurements involving parts of the given system 
and arbitrary nontrivial smooth function of two variables we can
reconstruct the dimensions of measures $\mu_1$, $\mu_2$ and $\mu_U$. 
From this we can obtain the dimension of interaction $d_q^{\sf
  int}$~(\ref{eq:dint}). Depending on the values of 
$D_q(\mu_1),D_q(\mu_2),D_q(\mu_U)$
and $d_q^{\sf int}$ we can find out if the systems
are coupled or not, and what is the direction of coupling.

\section{Classification of possible interaction schemes}
\label{sec:class}

Let us thus assume that we have two subsystems and the reconstructed
dimensions for them are $D_q(\mu_1)$ and $D_q(\mu_2)$. The
dimension of the whole system $D_q(\mu_U)$ is obtained from time
series $y(n)$ constructed through the procedure described in the
previous section. The dimension of interaction is calculated
from~(\ref{eq:dint}).  
The above discussion leads to a question which situations are
possible. There are four non-equivalent cases, which are conveniently
described by the following proposition 

{\bf Proposition 4.}{\em 
\begin{enumerate}
\item If $d_q^{\sf int}=0$, then $\mu_U=\mu_1\times \mu_2$ (the
  systems $U_1$ and $U_2$ do not interact);
\item If $D_q(\mu_1)=D_q(\mu_2)=d_q^{\sf int}$, then $\mu_U=
  \mu_1\equiv \mu_2\equiv \mu_V$ (the systems $U_1$ and $U_2$ are the
  same system or we have maximal coupling) ;
\item If $D_q(\mu_1)>D_q(\mu_2)=d_q^{\sf int}$, then $\mu_2=\mu_V$ and
  $\mu_1\equiv \mu_U$ (all variables of $U_2$ couple to some of the 
  degrees of freedom of $U_1$, or $U_2$ is the driver in the pair
  driver---response which is $U_1\equiv U$);
\item In all the other cases $D_q(\mu_1),D_q(\mu_2)>d_q^{\sf int}$,
  which means interaction or double control (two response systems
  driven by a common driver).
\end{enumerate}
}

Note that this proposition is to some extent opposite to the theorems
proposed in Section~\ref{sec:theory}. It can be shown for $q=1$
\cite{Wojcik00da}. We verify it numerically for particular systems for
$q=2$ in the next section. 

\begin{figure}[htbp]
  \begin{center}
    \leavevmode
    \begin{tabular}[c]{ccc}
      \includegraphics[scale=0.3]{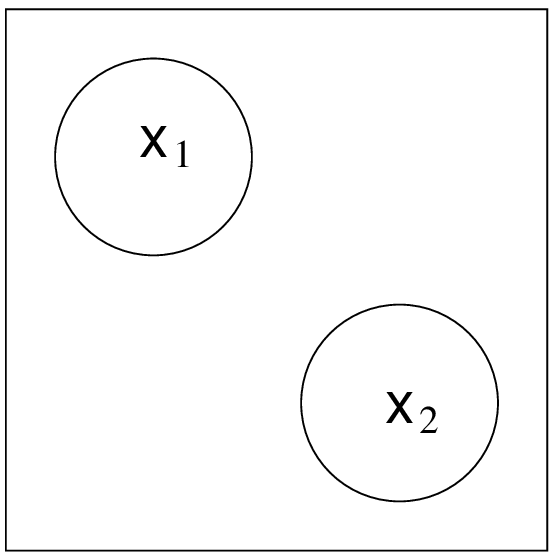} &
      \includegraphics[scale=0.3]{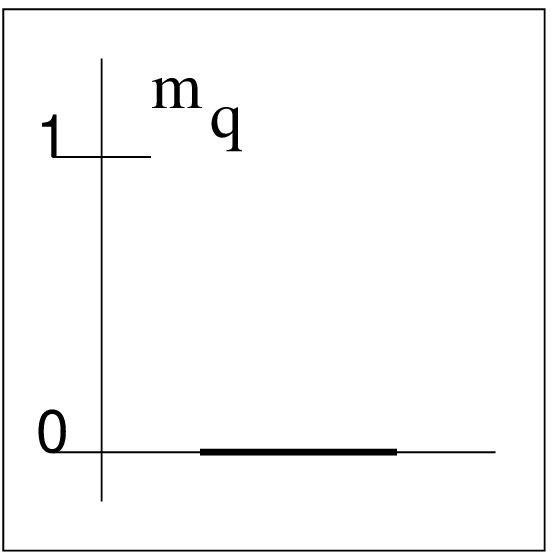} &
      $
      \begin{array}[b]{c}
        d_q^{\sf int}=0 \\
        \ \\
        m_1^q=m_2^q=m_U^q=0 \\
        \ \\ 
      \end{array}
      $
      \\
      \includegraphics[scale=0.3]{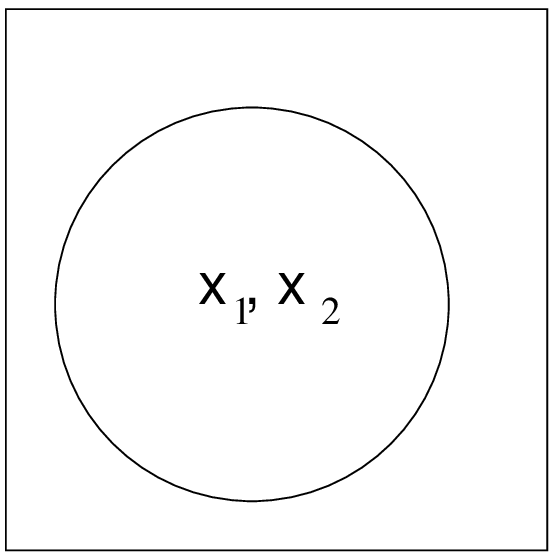} &
      \includegraphics[scale=0.3]{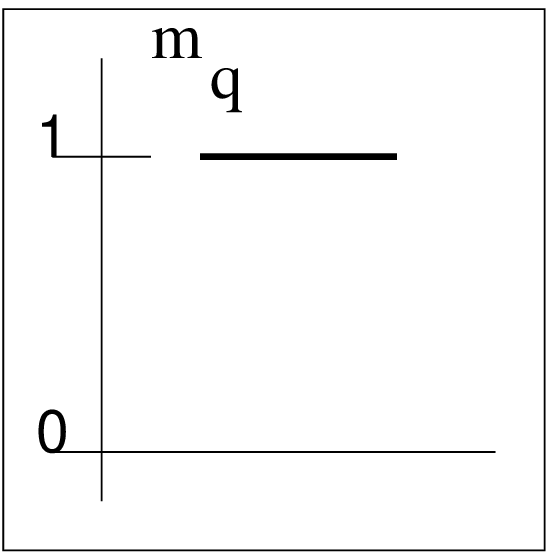} &
      $
      \begin{array}[b]{c}
        D_q(\mu_1)=D_q(\mu_2)=d_q^{\sf int}>0 \\
        \ \\
        m_1^q=m_2^q=m_U^q=1 \\
        \ \\
      \end{array}
      $  
      \\
      \includegraphics[scale=0.3]{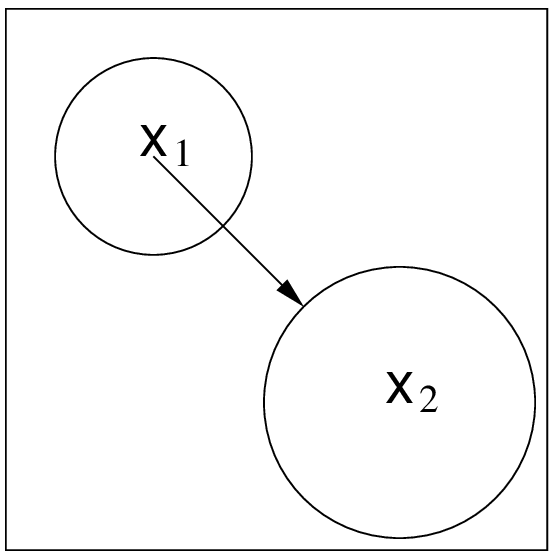} &
      \includegraphics[scale=0.3]{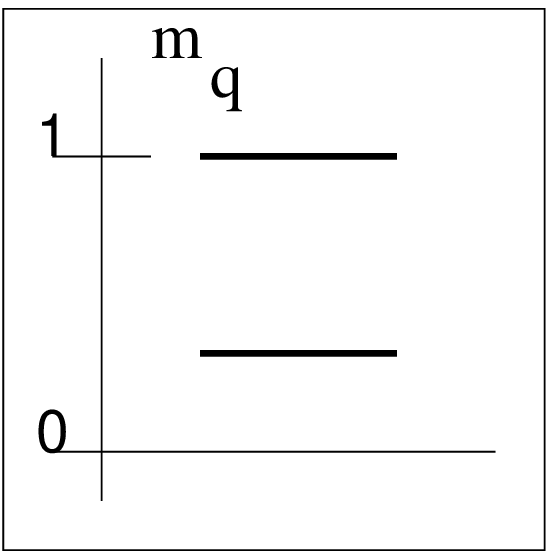} &
      $
      \begin{array}[b]{c}
        D_q(\mu_2)>D_q(\mu_1)=d_q^{\sf int}>0 \\
        \ \\
        1=m_2^q>m_1^q=m_U^q>0 \\
        \ \\ 
      \end{array}
      $
      \\
      \includegraphics[scale=0.3]{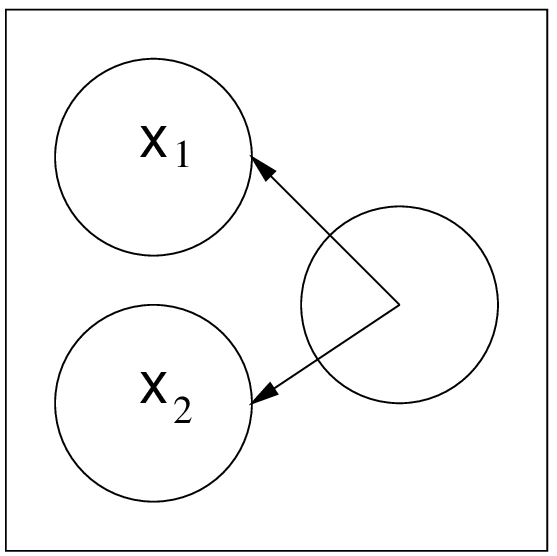} &
      \includegraphics[scale=0.3]{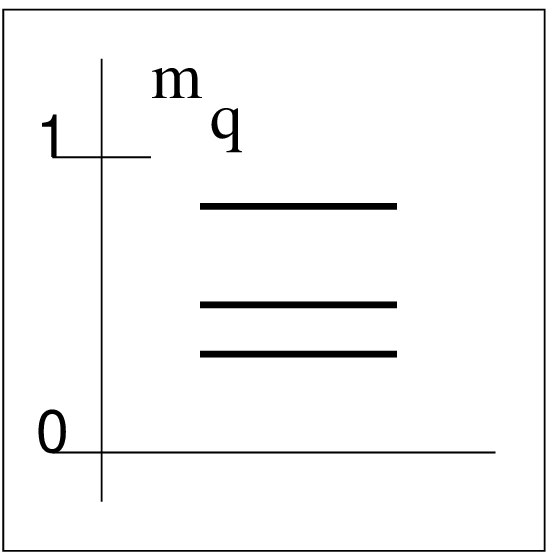} &
      $
      \begin{array}[b]{c}
        D_q(\mu_1),D_q(\mu_2)>d_q^{\sf int}>0 \\
        \ \\ 
        1>m_1^q\geq m_2^q>m_U^q>0 \\
        \ \\ 
      \end{array}
      $
    \end{tabular}
    \caption{Classification of possible interaction schemes. The first
      column shows symbolically the relative position in the phase
      space of the subsystems in which
      we measure the time series $x_1$ and $x_2$. An arrow from
      one system to another means the future states of the second
      system depend on the current states of both. The second column
      shows the values of normalized dimensions $m_1^q$, $m_2^q$ and
      $m_U^q$ in each of the cases.}
    \label{fig:class}
  \end{center}
\end{figure}

It is convenient sometimes to use $m_1^q, m_2^q, m_U^q$
(\ref{eq:normdim}). We can write the above classification in this case
as follows
\begin{enumerate}
\item $m_1^q=m_2^q=m_U^q=0$ means no interaction;  
\item $1=m_1^q=m_2^q=m_U^q$ means maximal coupling: $\mu_1\equiv
  \mu_2\equiv \mu_V$; 
\item $1=m_1^q>m_2^q=m_U^q>0$ means coupling of all the degrees of
  freedom of $U_2\equiv V$ to some variables of $U_1$;
\item $1>m_1^q\geq m_2^q>m_U^q>0$ means interaction or double control (two
  systems driven by a common driver); 
\end{enumerate}
All the four cases are presented symbolically on
Figure~\ref{fig:class}. 

The examples considered in the previous section can be easily
identified as particular cases of this classification. Namely
example~I represents case~1, example~II represents case~4, example~III
can represent cases 2, 3 or 4, the last example represents cases 1 or
4 depending on whether the signals analyzed come from systems coupled
to the same driver or not. 

\section{Numerical results}
\label{sec:results}

Below we shall present some applications of our method to analysis of
numerical results for several paradigmatic systems (coupled H\'enon
maps and logistic maps). 

Throughout this section we will use $d_2^{\sf int}$. 
The dimensions presented in the pictures are always $D_2$ calculated
with the help of {\tt d2} program from TISEAN package
\cite{hegger99da} with an algorithm which is an extension of   
algorithms published previously 
\cite{Grassberger83c,Grassberger83d,Kantz93c} which 
improves speed of computation \cite{hegger99da}. In every case we used
$10^5$ points with one exception described in the text. The functions
$Y$ used to calculate the dimension of the whole system (cf. previous
sections) were $x+y$, $x\cdot y$, $\sin(x)\cos(y)$, $x\exp(y)$,
$2x-y$. To estimate the dimension we used Takens-Theiler estimator
\cite{Kantz97c,hegger99da,Takens85a,Theiler88} {\tt c2t} and {\tt c2d}
smoothed output from {\tt d2}.  

\begin{figure}[htbp]
  \begin{center}
      \includegraphics[scale=0.5]{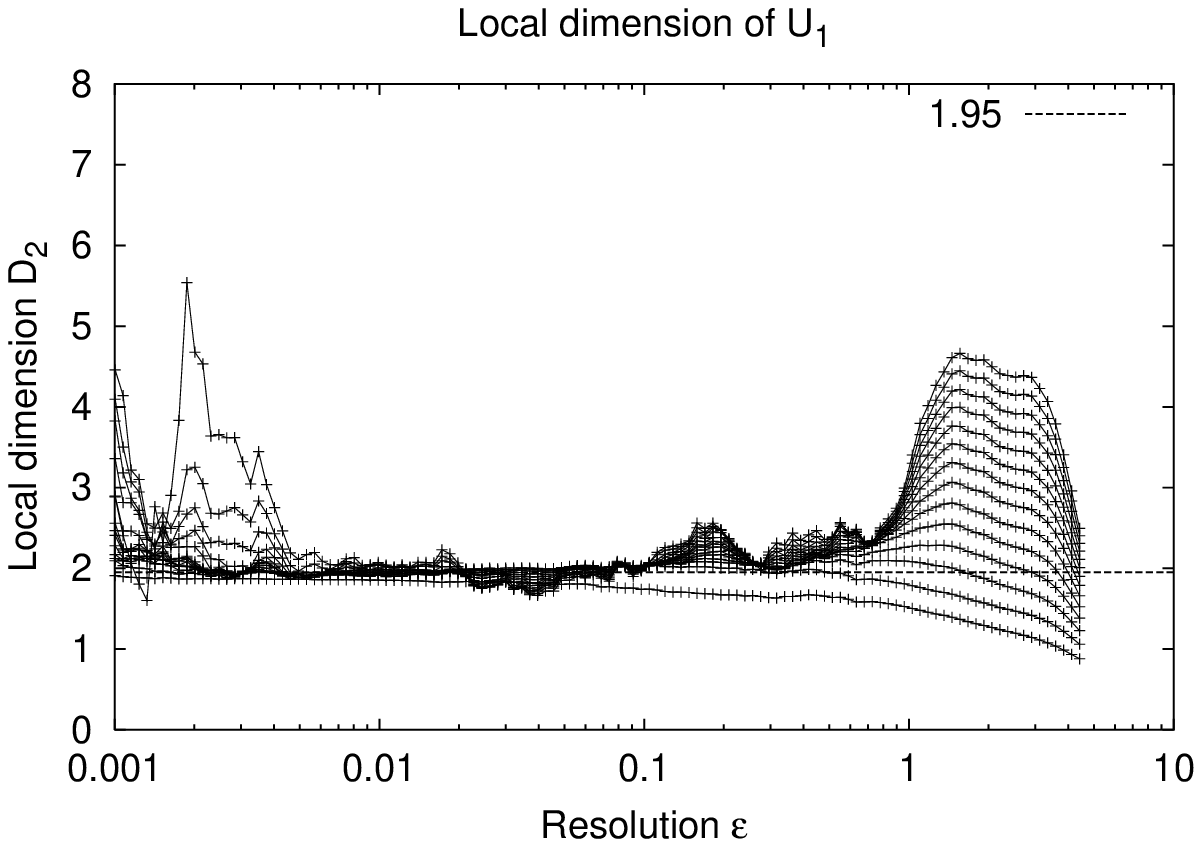} \\
      \includegraphics[scale=0.5]{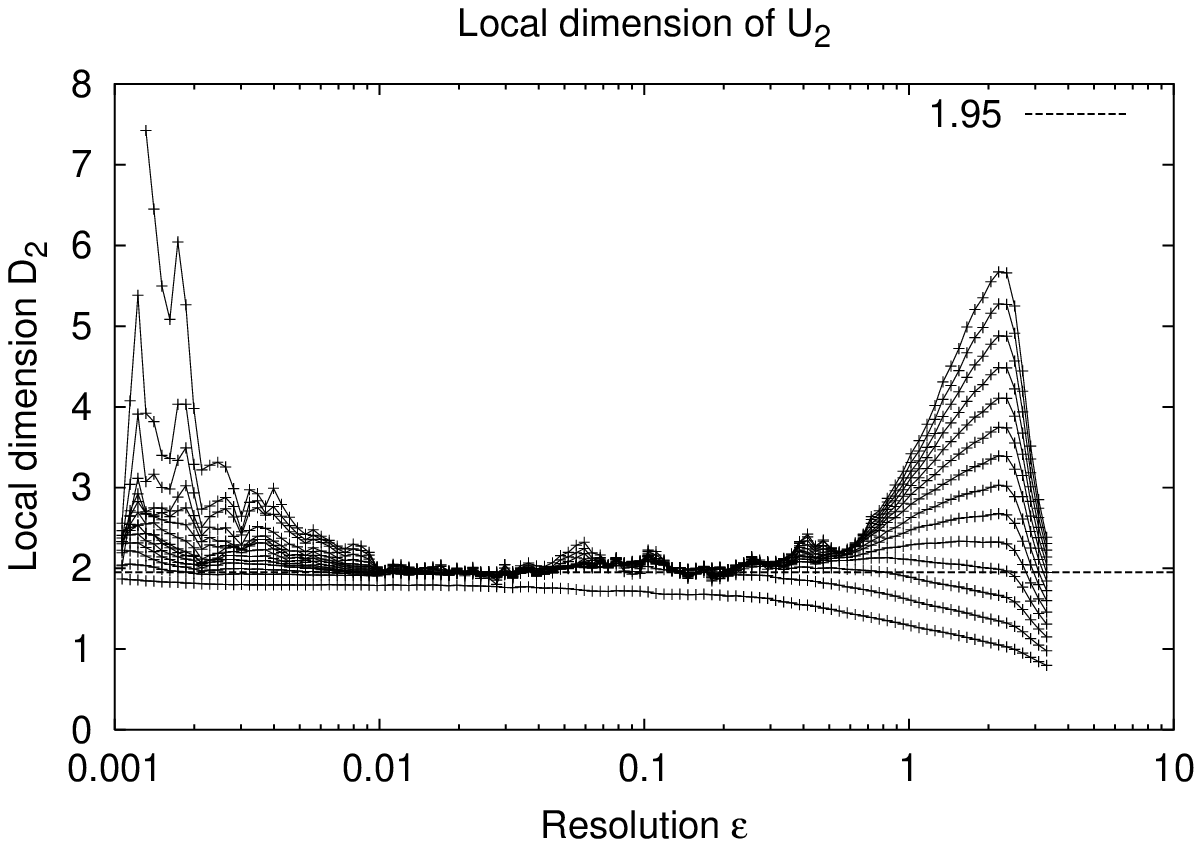} \\    
      \includegraphics[scale=0.5]{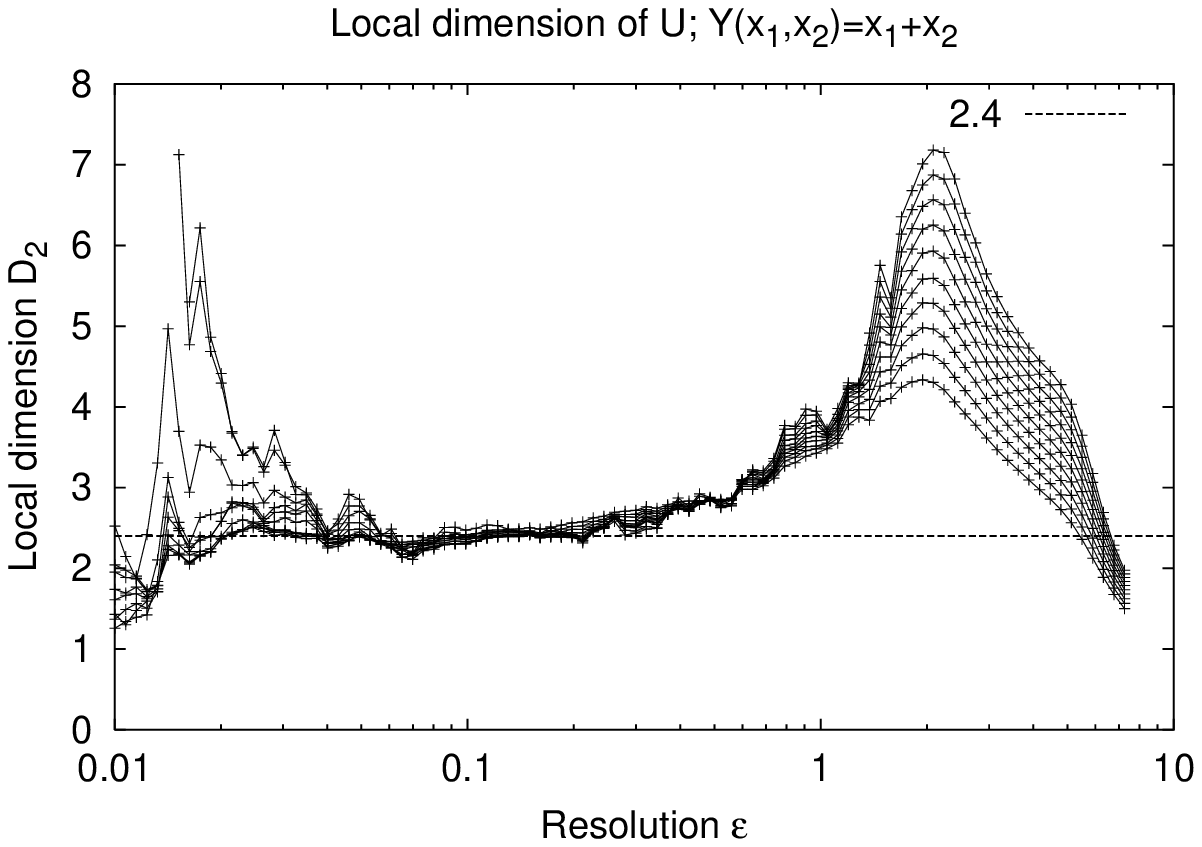} \\
      \includegraphics[scale=0.5]{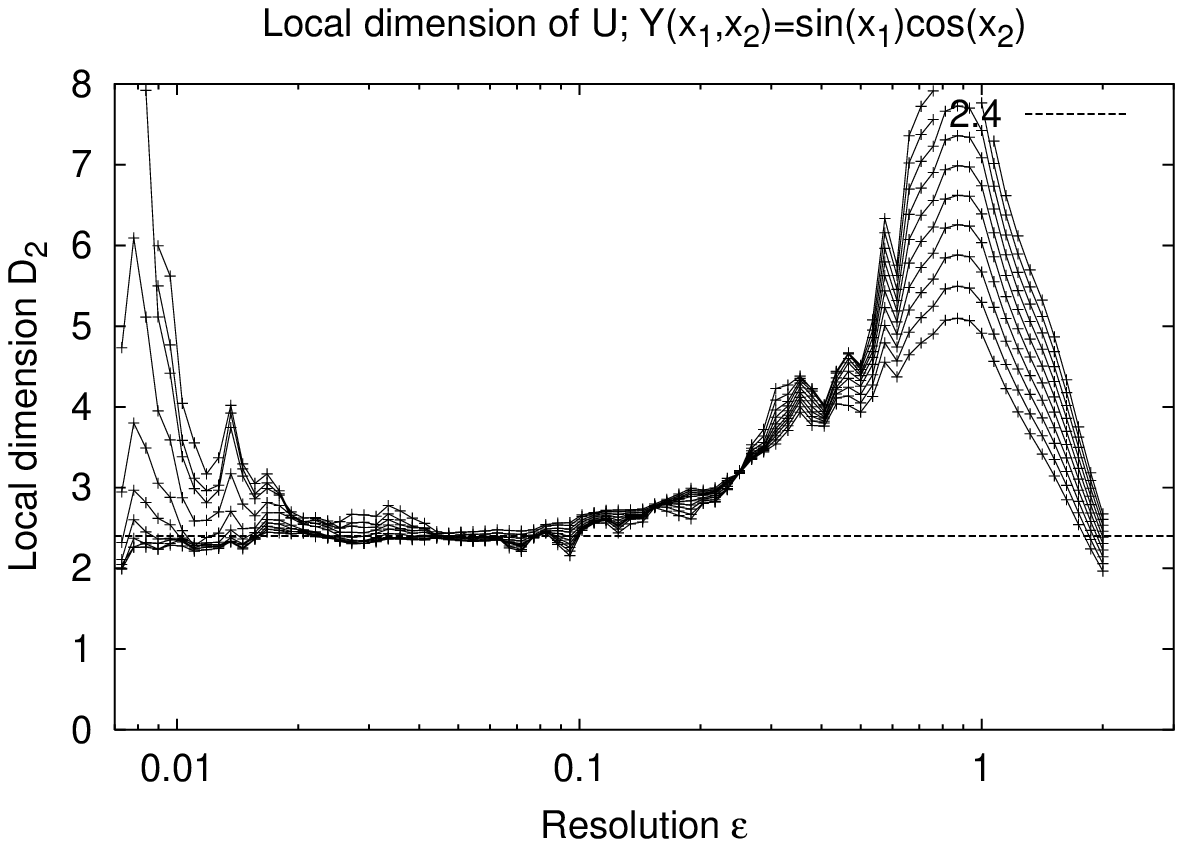} 
    \caption{Takens estimator of correlation dimension. Data shown
      come from two different H\'enon systems driven by the third
      (eq. \ref{coupledthreehenons} with parameters $B_1=0.3$,
      $C_1=0.5$, $B_2=0.1$, $C_2=0.6$.). Correlation dimension
      estimated from the pictures is 1.95 for both of the subsystems,
      and 2.4 for the whole system. We show two plots out of five used
      to estimate the last number. Dimension of interaction in this
      case is $1.95+1.95-2.4=1.5>1.22$, which suggests partial
      synchronization of the two response systems with the driver.}
    \label{fig:resol}
  \end{center}
\end{figure}

Typical behavior of local dimension $d \log C(\varepsilon)/ d \log \varepsilon$
as a function of resolution $\varepsilon$ is shown in
figure~\ref{fig:resol}. 

\subsection{Two H\'enon maps}

Consider a system $U$ consisting of two H\'enon maps \cite{Hnon76}
coupled as follows 
\cite{Schiff96}:   
\begin{eqnarray}
& K &
{\left\{
\begin{array}{rcl}
  x_{i+1} & = & 1.4-x_i^2+0.3 y_i,\\
  y_{i+1} & = & x_i,
\end{array}
\right.}\nonumber \\
& L & 
{\left\{
\begin{array}{rcl}
  u_{i+1} & = & 1.4-(C x_i+(1-C)u_i)u_i+B v_i,\\
  v_{i+1} & = & u_i,
\end{array}
\right.} \label{coupledtwohenons}
\end{eqnarray}
Thus H\'enon system $K$ drives system $L$. The coupling is introduced
through variable $u$. We consider the case of coupled
identical systems ($B=0.3$) and non-identical coupled systems
($B=0.1$). Parameter $C$ measures the strength of interaction.

Suppose the variables accessible experimentally are $x_n$ and $u_n$. 
What can be said in this case about the interaction between systems
$K$ and $L$? 

Certainly, for $C=0$ the systems $K$ and $L$ do not interact (case
1. in our classification), therefore
$D_q(\mu_U)=D_q(\mu_K)+D_q(\mu_L)$ and $d_q^{\sf int}=0$. On the other 
hand, for positive $C$ the influence of $x$ should reflect in the
behavior of $u$. From Theorem 3 we expect $D_q^{\sf int}=
D_q(\mu_K)$ (case~3.). One can also expect for $C$ raising slightly  
above $0$, $D_q(\mu_U)$ not to change much, while
$D_q(\mu_L)$ should jump from its value at $0$ to the value of
$D_q(\mu_U)$ at $c=0$.
\begin{figure}[htbp]
  \begin{center}
    \begin{tabular}[c]{c}
      (a) \hfill\ \\
      \includegraphics[scale=0.6]{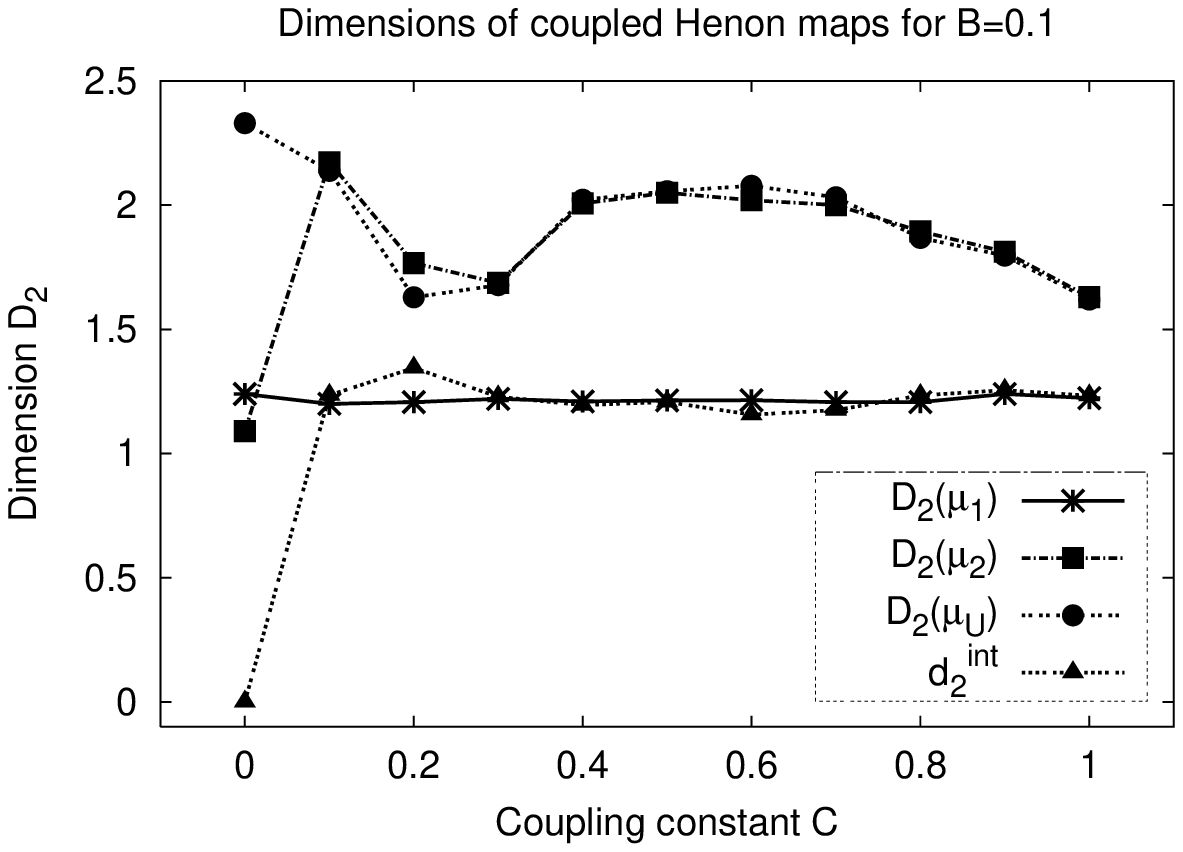}\\
      (b) \hfill\ \\    
      \includegraphics[scale=0.6]{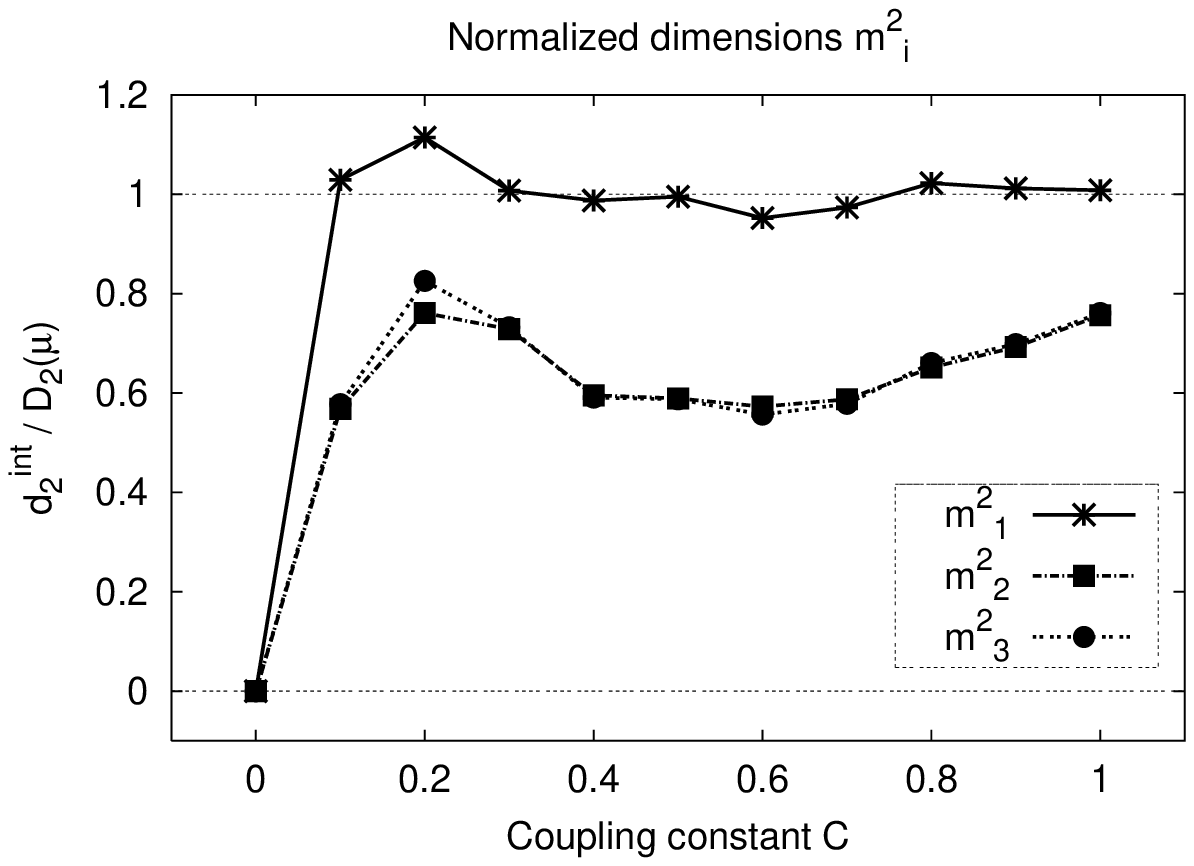}
    \end{tabular}
    \caption{a) Dimensions $D_2(\mu_1), D_2(\mu_2),D_2(\mu_U)$ and
      $d_2^{\sf int}$ of one-way coupled non-identical H\'enon
      maps~(\ref{coupledtwohenons}) $B=0.1$. b) Normalized
      dimensions $m_1^2$, $m_2^2$ and $m_U^2$ for the same systems.}  
    \label{fig:onewayhenon1}
  \end{center}
\end{figure}
\begin{figure}[htbp]
  \begin{center}
    \begin{tabular}[c]{c}
      (a) \hfill\ \\
      \includegraphics[scale=0.6]{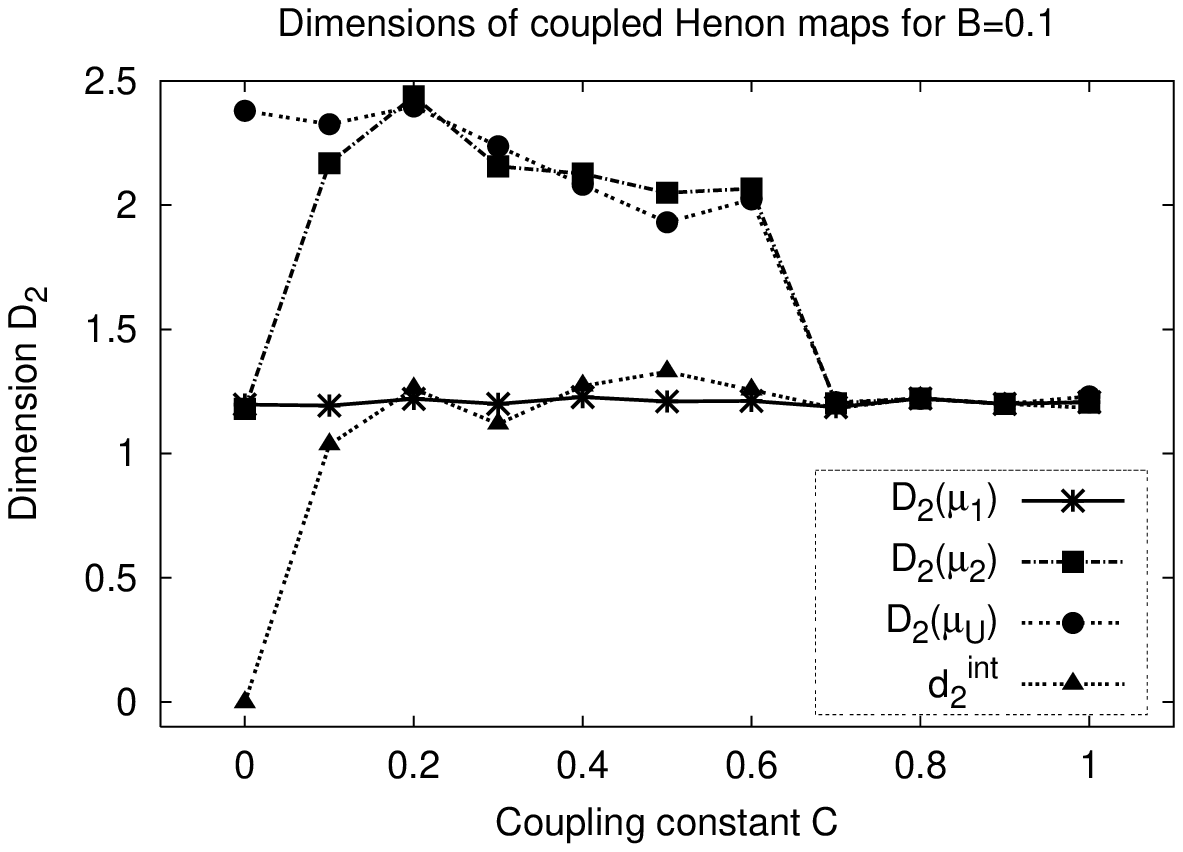}\\
      (b) \hfill\ \\ 
      \includegraphics[scale=0.6]{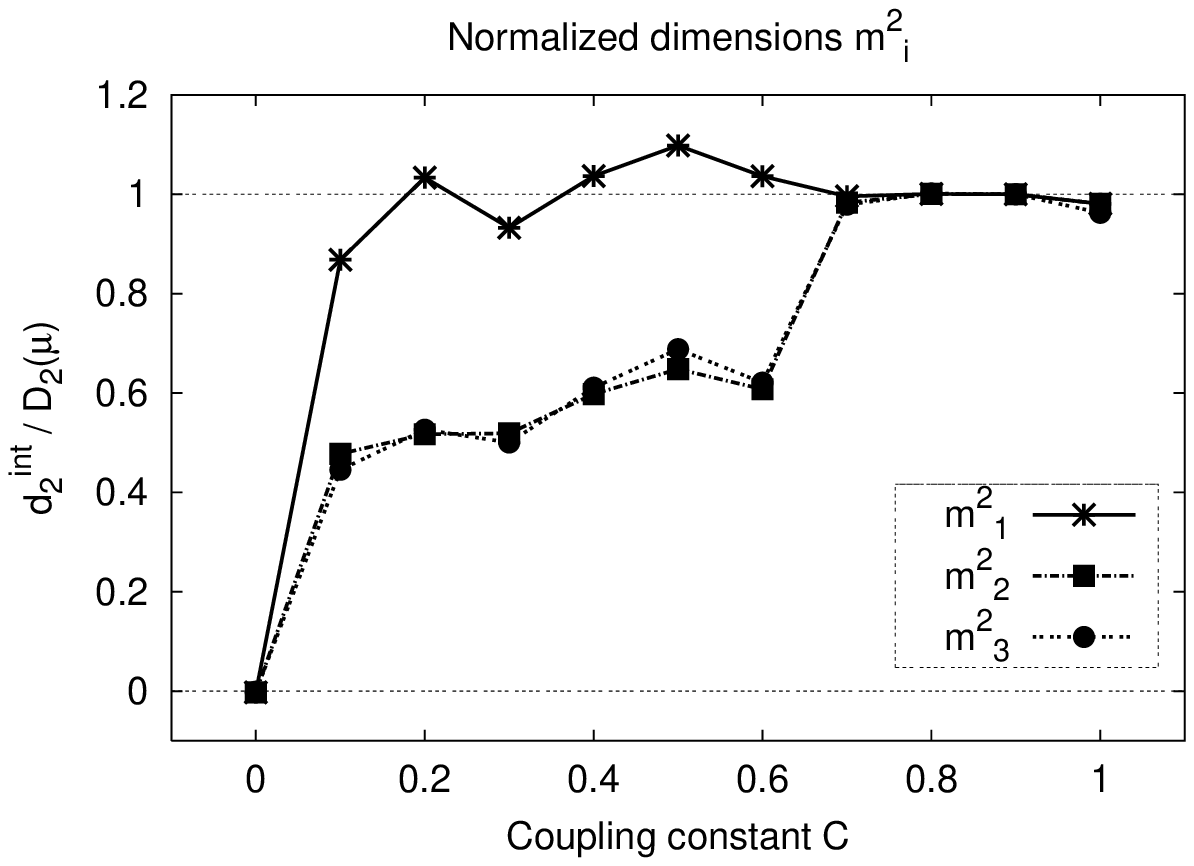}
    \end{tabular}
    \caption{a) Dimensions $D_2(\mu_1), D_2(\mu_2),D_2(\mu_U)$ and
      $d_2^{\sf int}$ of one-way coupled identical H\'enon
      maps~(\ref{coupledtwohenons}) $B=0.3$. b) Normalized
      dimensions $m_1^2$, $m_2^2$ and $m_U^2$ for the same systems.}  
    \label{fig:onewayhenon2}
  \end{center}
\end{figure}                              

This behavior can indeed be seen in figure~\ref{fig:onewayhenon1}a
for non-identical H\'enon systems 
($B=0.1$) and in~\ref{fig:onewayhenon2}a for identical systems
($B=0.3$). The synchronization of $x$ and $u$\cite{Pecora90} visible
for $C\geq 0.7$ (case~2.) can be discovered much simpler, namely if one plots
several consecutive values of $x_n-u_n$ (100, say) versus coupling,
for these particular values all of the points fall on 0 (cf. figure~7
of \cite{Schiff96}).  

Looking at the normalized dimensions 
(fig.~\ref{fig:onewayhenon1}b and~\ref{fig:onewayhenon2}b) we easily
identify lack of coupling for $C=0$ ($m_1=m_2=m_3=0$), case 4. (maximal
coupling) for $B=0.3$ and $C\geq 7$ ($m_1=m_2=m_3=1$), and case
3. in all the other cases.

The drop-down of the dimension at 0.7 for identical systems is
connected with the full synchronization of the systems. The equations
(\ref{coupledtwohenons}) admit solutions symmetric in $x$ and $u$
($x_n-u_n=0$), which at this region become stable and the whole
probability measure gets localized on a lower-dimensional
manifold. For more details cf. \cite{Schiff96}.

\subsection{Three H\'enon maps}

Consider now the system $U$ consisting of three H\'enon
maps \cite{Hnon76} coupled as follows \cite{Schiff96}:   
\begin{eqnarray}
& K &
{\left\{
\begin{array}{rcl}
  x_{i+1} & = & 1.4-x_i^2+0.3 y_i,\\
  y_{i+1} & = & x_i,
\end{array}
\right.}\nonumber \\
& L & 
{\left\{
\begin{array}{rcl}
  u_{i+1} & = & 1.4-(C_1 x_i+(1-C_1)u_i)u_i+B_1 v_i,\\
  v_{i+1} & = & u_i,
\end{array}
\right.} \label{coupledthreehenons}\\ 
& M & 
{\left\{
\begin{array}{rcl}
  w_{i+1} & = & 1.4-(C_2 x_i+(1-C_2)w_i)w_i+B_2 z_i,\\
  z_{i+1} & = & w_i.
\end{array}
\right.} \nonumber
\end{eqnarray}
Thus H\'enon system $K$ drives systems $L$ and $M$. The coupling is
introduced through variables $u$ and $w$. Parameters $C_1,C_2$ measure
the strength of interaction. 

Suppose the measurements on $(K,L,M)$ yield variables $u$ and $w$.
What can be said in this case about the interaction between the
systems $L$ and $M$?

For $C_1=C_2=0$ neither $L$ nor $M$ systems feel the influence of
$K$. They also do not interact (case~1.). When one of $C_i$ grows,
the influence of $K$ is immediately mirrored in the rise of the
dimension of $\mu_L$ or $\mu_M$. For both $C_i>0$ the systems $L$
and $M$ interact (case~2.), and the part responsible for interaction
is $K$. Thus the dimension of the common part is constant and equal
to 1.22 in our case.

\begin{figure}[htbp]
    \begin{center}
      \begin{tabular}[c]{c}
        (a) \hfill\ \\
        \includegraphics[scale=0.6]{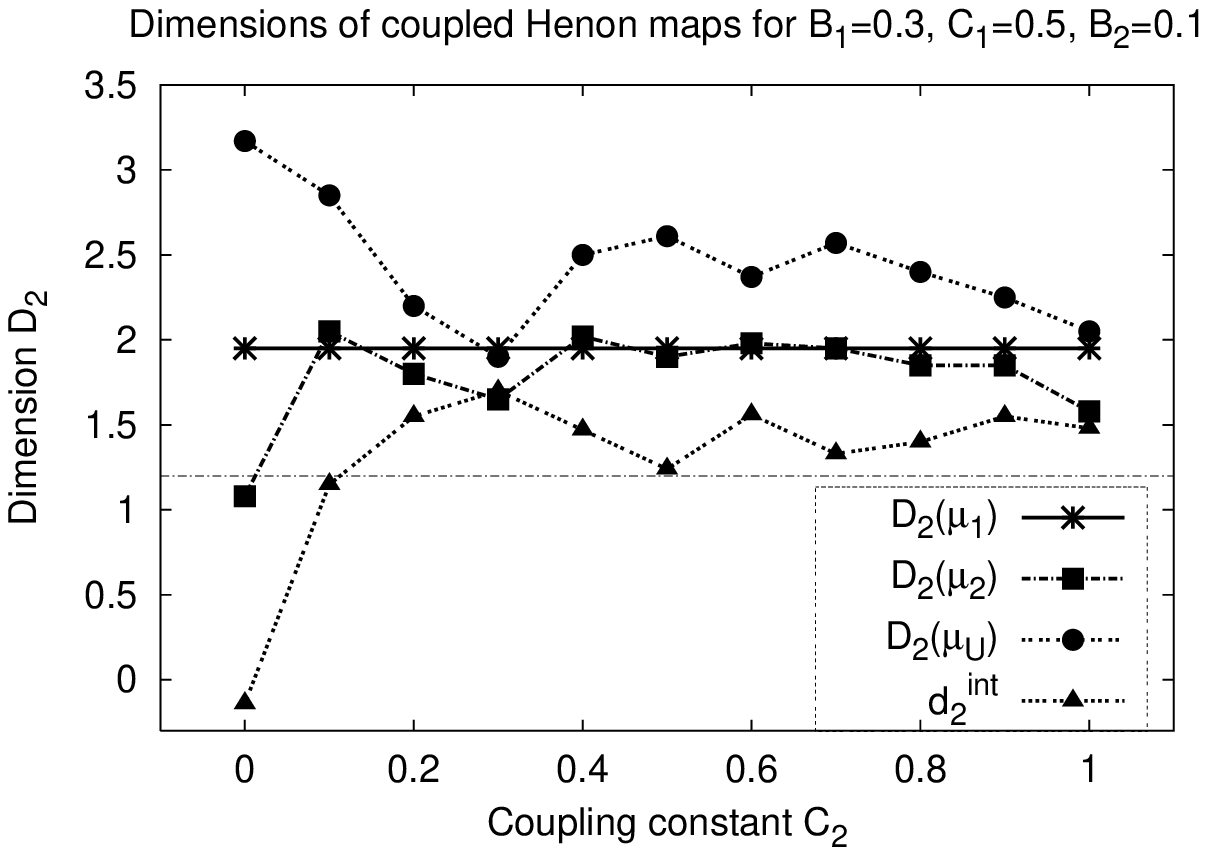}\\
        (b) \hfill\ \\
        \includegraphics[scale=0.6]{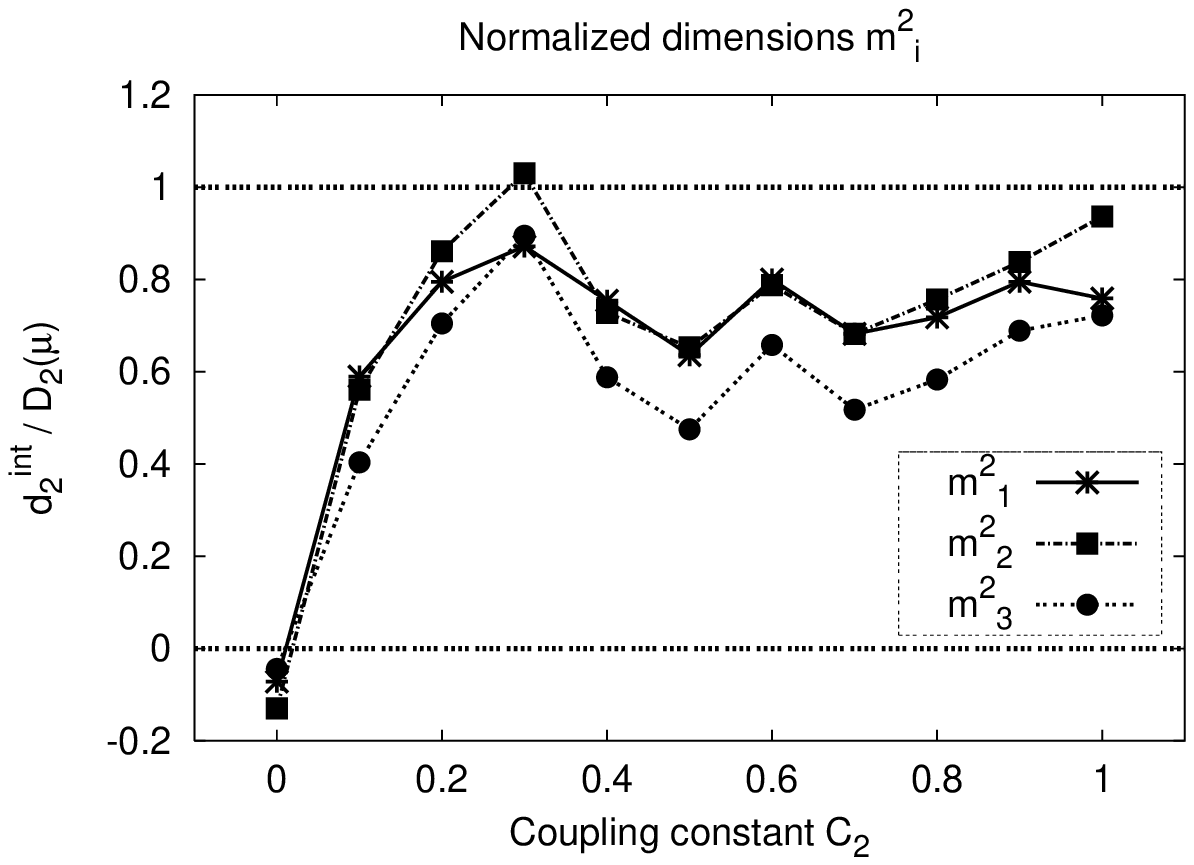}
      \end{tabular}
      \caption{a) Dimensions $D_2(\mu_1), D_2(\mu_2),D_2(\mu_U)$ and
        $d_2^{\sf int}$ of two-way coupled H\'enon
        maps~(\ref{coupledthreehenons}) with different response
        systems ($C_1=0.5, 
        B_1=0.3, B_2=0.1$). Additional line at 1.2 in the upper figure
        stands for the dimension of the attractor of H\'enon system $K$.
        b) Normalized dimensions $m_1^2$, $m_2^2$ and $m_U^2$ for the
        same systems.}  
      \label{fig:twowayhenon1}
    \end{center}
  \end{figure}

\begin{figure}[htbp]
    \begin{center}
      \begin{tabular}[c]{c}
        (a) \hfill\ \\
        \includegraphics[scale=0.6]{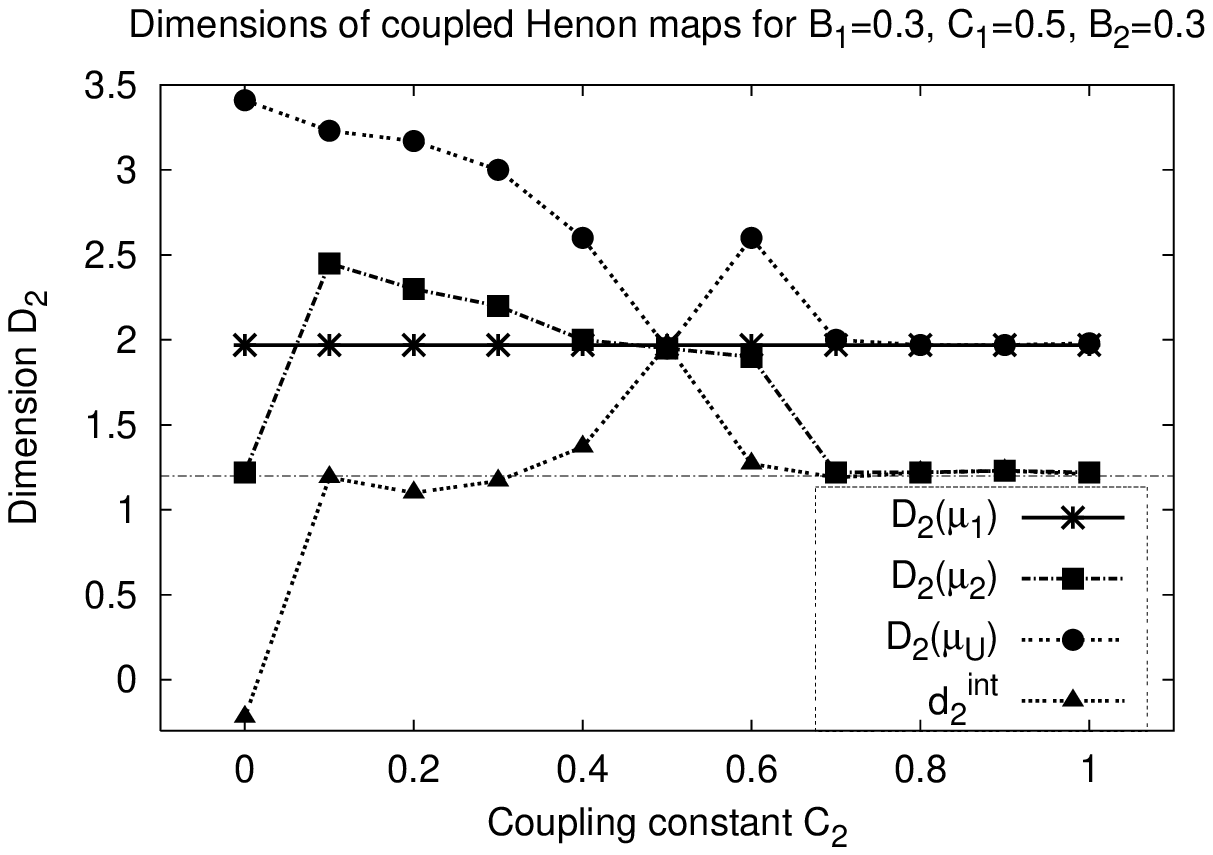}\\
        (b) \hfill\ \\
        \includegraphics[scale=0.6]{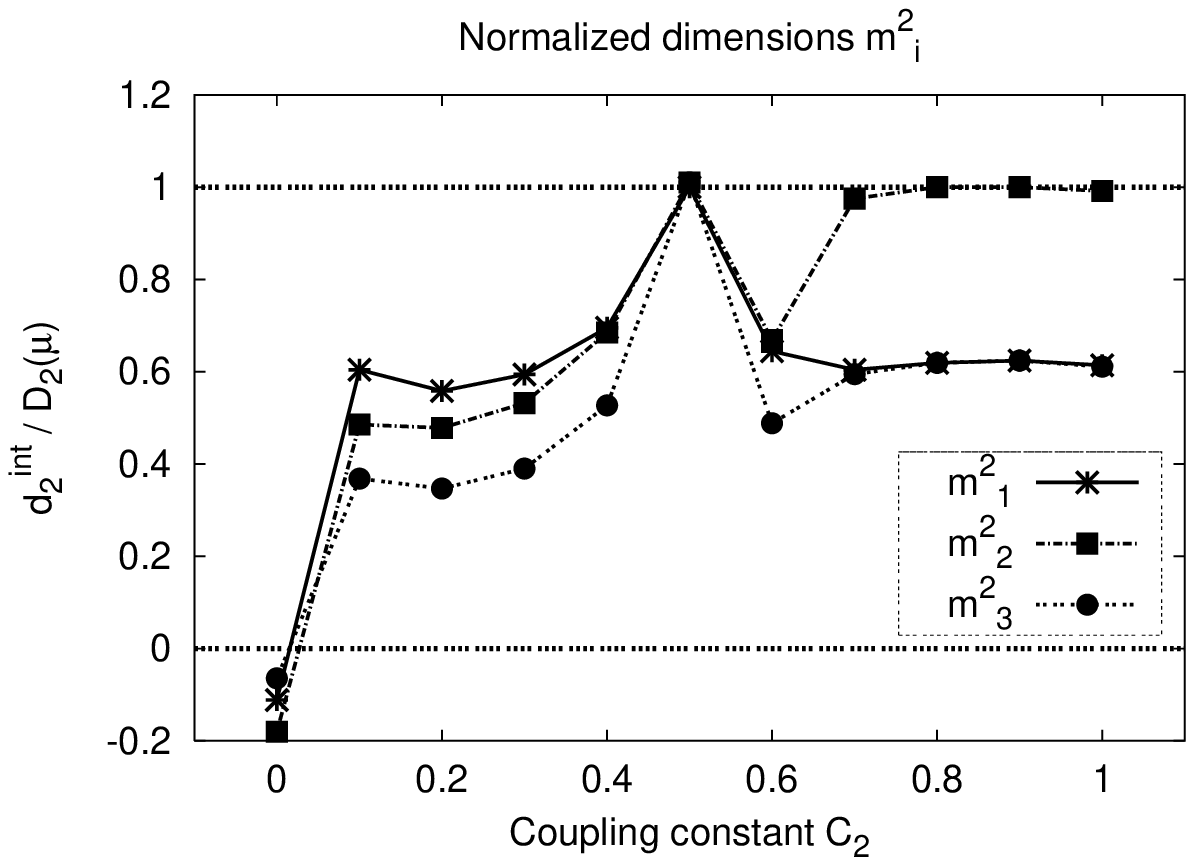}
      \end{tabular}
      \caption{a) Dimensions $D_2(\mu_1), D_2(\mu_2),D_2(\mu_U)$ and
        $d_2^{\sf int}$ of two-way coupled H\'enon
        maps~(\ref{coupledthreehenons}) with identical response
        systems ($C_1=0.5, 
        B_1=0.3, B_2=0.3$). Additional line at 1.2 in the upper figure
        stands for the dimension of the attractor of H\'enon system $K$.
        b) Normalized dimensions $m_1^2$, $m_2^2$ and $m_U^2$ for the
        same systems.}  
      \label{fig:twowayhenon2}
    \end{center}
  \end{figure}

We show this behavior in figures~\ref{fig:twowayhenon1}a for
different parameters ($B_1=0.3, B_2=0.1$) and~\ref{fig:twowayhenon2}a 
for the same parameters ($B_1=0.3, B_2=0.3$). In both cases
$C_1=0.5$ and $C_2$ is varied. In both figures one can clearly see
the jump of the dimension of interaction from 0 to values equal to
or greater than 1.22, the dimension of the attractor of $K$.

Figure~\ref{fig:twowayhenon2} is particularly interesting, since one
can apparently identify all the four cases from our
classification. For $C_2=0$ we have non-interacting systems, for
$C_2\in[0.2,0.4]$ and $C_2=0.6$ we have case 2.

For $C_2=0.5$ the two H\'enon systems $L$ and $M$ become
identical. Since at this value of coupling constant they are in
general synchrony with the driver, which means their asymptotic
states are independent of their initial states, and depend only on
the present state of the driver, it follows that $u_n=w_n$.

For $C_2\geq 0.7$ the system $M$ fully synchronizes with $K$, which
leads to the collapse of the probability measure in $K,M$ space on
the diagonal (compare the discussion in the previous subsection).

\subsection{Logistic maps}

Let $f_\alpha(x):=\alpha x(1-x)$.
Consider a system consisting of four uncoupled logistic maps
\[
x^i_{n+1}=f_{\alpha_i}(x^i_n),
\]
where $\alpha_1=3.7$, $\alpha_2=3.8$, $\alpha_3=3.9$ and $\alpha_4=4$. 
Suppose the only variables available experimentally are\footnote{The
  coupling functions $F^{i,j}$ were chosen randomly out of $x+y$, $x\cdot
  y$, $\sin(x)\cos(y)$, $x\exp(y)$, $2x-y$.}
$Y^{i,j}(n)=F^{i,j}(x^i_n,x^j_n),\,i< j$ . Given two 
randomly chosen time series $Y^{i,j}(n),Y^{k,l}(n)$ we want to know if they share
some degrees of freedom or not (if they ``interact'' or not). If $i$ or
$j$ is equal to $k$ or $l$, there are only three active degrees of freedom
in the compound system. Otherwise there are four. 

Estimated correlation dimensions for several cases are collected in
Table~\ref{fig:tabdim1}. In every case we used time series $10^5$
points long except for the last one, for which $10^6$ points were
used. The estimation error was roughly 2\% except for the last case
for which it was about 5-10\%\footnote{We believe there are two reasons for
this. One is higher dimensionality of the system in the last case
(four uncupled logistic maps), the other is worse ergodicity in the
phase space because the maps are uncoupled. Note that our procedure
consists of two parts: first we make the embedding, then we calculate
the dimensions. Each of the two can introduce errors. The number
expected in the last case is the sum of the first four numbers, namely
3.87}.

\begin{figure}[htbp]
\[
\begin{array}[c]{cccc}
\hline
{\rm series}\, x(n) & D(\mu_{x(n)}) \\
\hline
x_1 & 0.96\\
x_2 & 0.95\\
x_3 & 0.97\\
x_4 & 0.99\\
Y^{1,2} & 1.88\\
Y^{1,3} & 1.94\\
Y^{1,4} & 1.95\\
Y^{2,3} & 1.89\\
Y^{2,4} & 1.94\\
Y^{3,4} & 1.93\\
f(Y^{1,2},Y^{1,3}) & 2.88\\
f(Y^{1,2},Y^{3,4}) & 3.8\\ 
\hline
\end{array}
\]
\caption{Estimated correlation dimension for uncoupled logistic
  maps. The estimation error is roughly 2\% except for the last number
  for which it is about 5-10\%.} 
\label{fig:tabdim1}
\end{figure}

Consider now two symmetrically coupled logistic maps
\begin{equation}
\left\{
  \begin{array}[c]{rcl}
    x_{n+1} & = & f_\alpha (\tilde{x}_n),\\
    y_{n+1} & = & f_\beta (\tilde{y}_n),
  \end{array}
\right.
\quad
\mbox{\rm where}
\quad
\left\{
\begin{array}[c]{rcl}
  \tilde{x}_n & = & \frac{x_n+c y_n}{1+c},\\
  \tilde{y}_n & = & \frac{y_n+c x_n}{1+c},
\end{array}
\right.
\label{eq:couple}
\end{equation}
and parameter $c\in[0,1]$ measures the coupling. This is slightly
different from couplings discussed previously in the literature
(e.g. \cite{Pikovsky91,Zochowski97,Schuster98da}). The maps are uncoupled for
$c=0$. For $c=1$ (the strongest coupling) if we set
$z_n:=\tilde{x}_n=\tilde{y}_n$, we have
$x_n=\frac{2\alpha}{\alpha+\beta}z_n$, 
$y_n=\frac{2\beta}{\alpha+\beta}z_n$, and
$z_{n+1}=f_{\frac{\alpha+\beta}{2}}(z_n)$. Therefore dynamics is 
one-dimensional. Case $c>1$ is equivalent to $c'=1/c$.
\begin{figure}[htbp]
  \begin{center}
    \begin{tabular}[c]{c}
      (a) \hfill\ \\
      \includegraphics[scale=0.6]{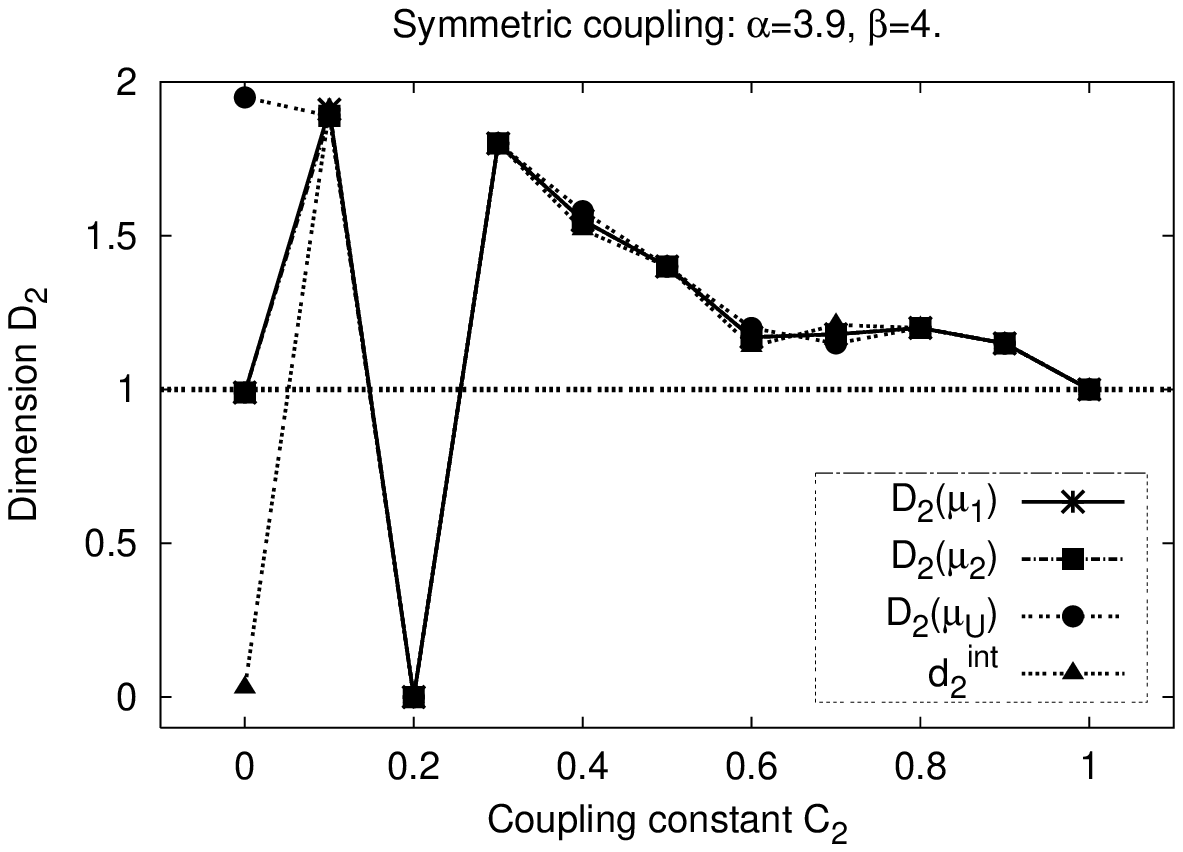}\\
      (b) \hfill\ \\
      \includegraphics[scale=0.6]{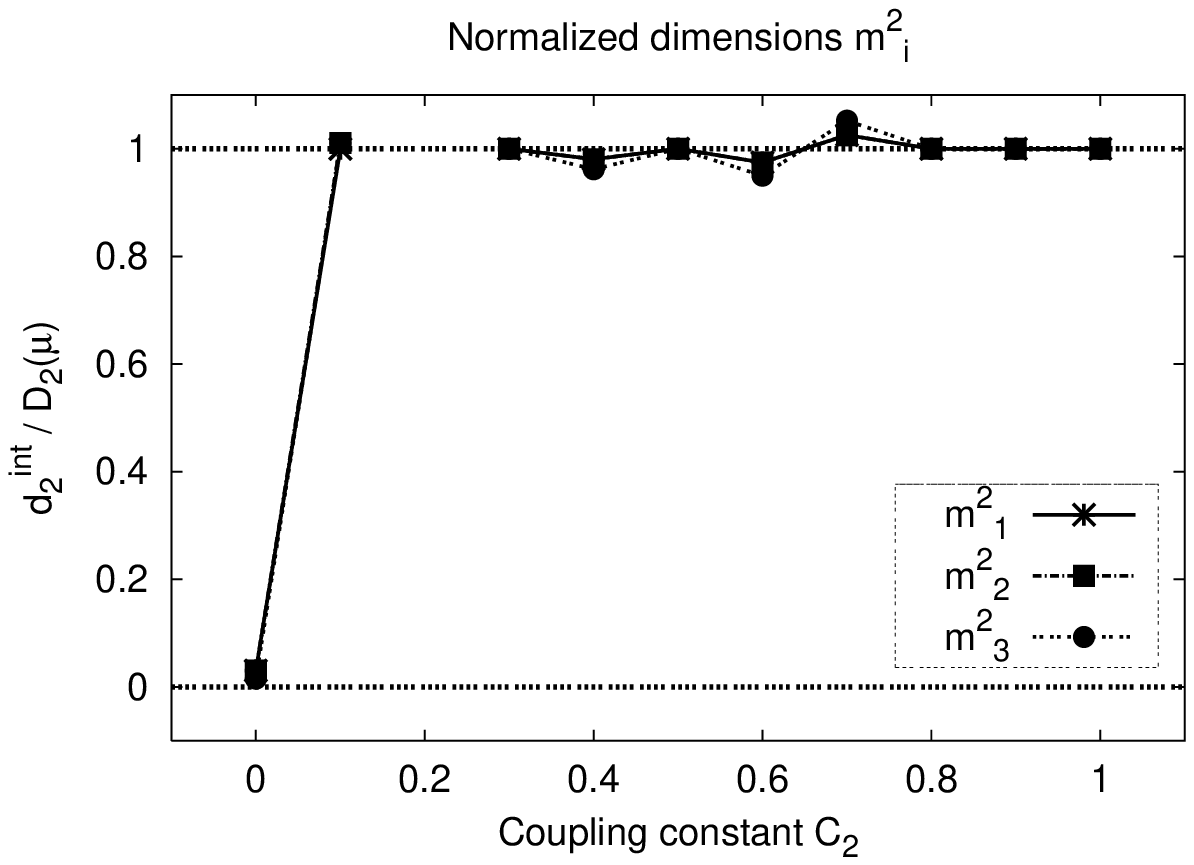}
    \end{tabular}
    \caption{a) Dimensions $D_2(\mu_1), D_2(\mu_2),D_2(\mu_U)$ and
      $d_2^{\sf int}$ of symmetrically coupled logistic maps
      (\ref{eq:couple}). b) Normalized
      dimensions $m_1^2$, $m_2^2$ and $m_U^2$ for the same systems.}
    \label{fig:symlogi}
  \end{center}
\end{figure}      
  
\begin{figure}[htbp]
  \begin{center}
    \leavevmode
    \begin{tabular}[c]{c} \hskip -0.5cm
      \includegraphics[scale=0.5]{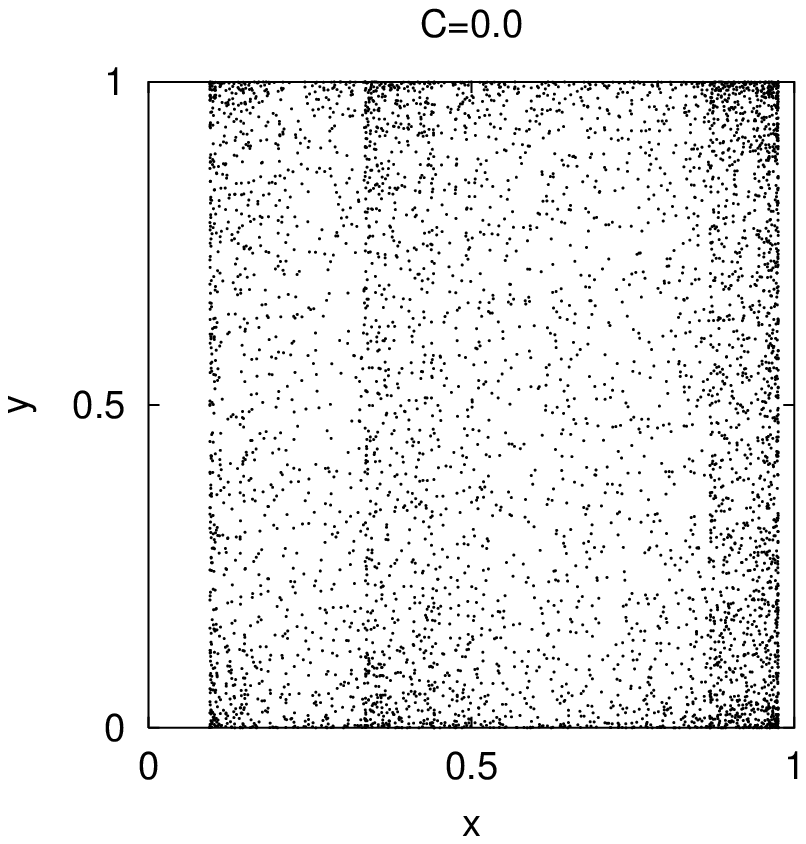}\hskip -2cm
      \includegraphics[scale=0.5]{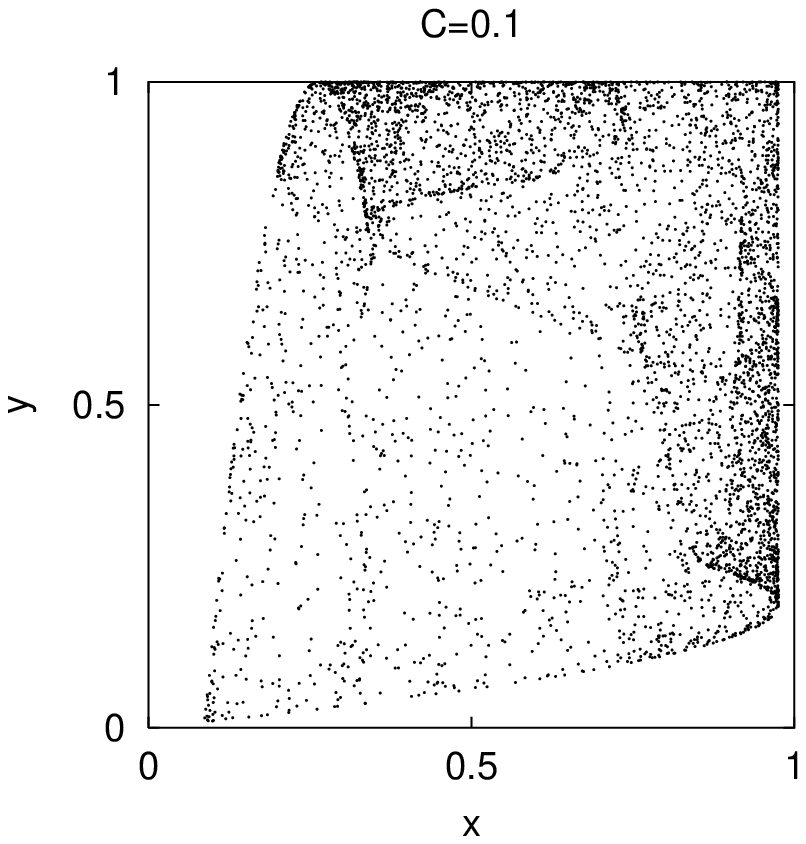}\\
     \hskip -0.5cm \includegraphics[scale=0.5]{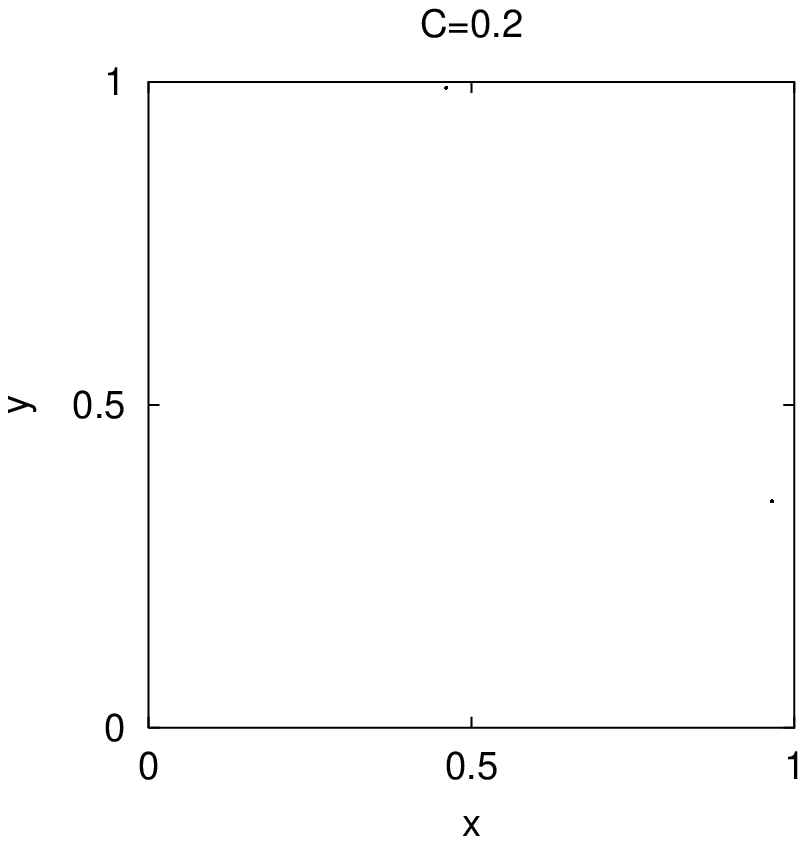}\hskip -2cm
      \includegraphics[scale=0.5]{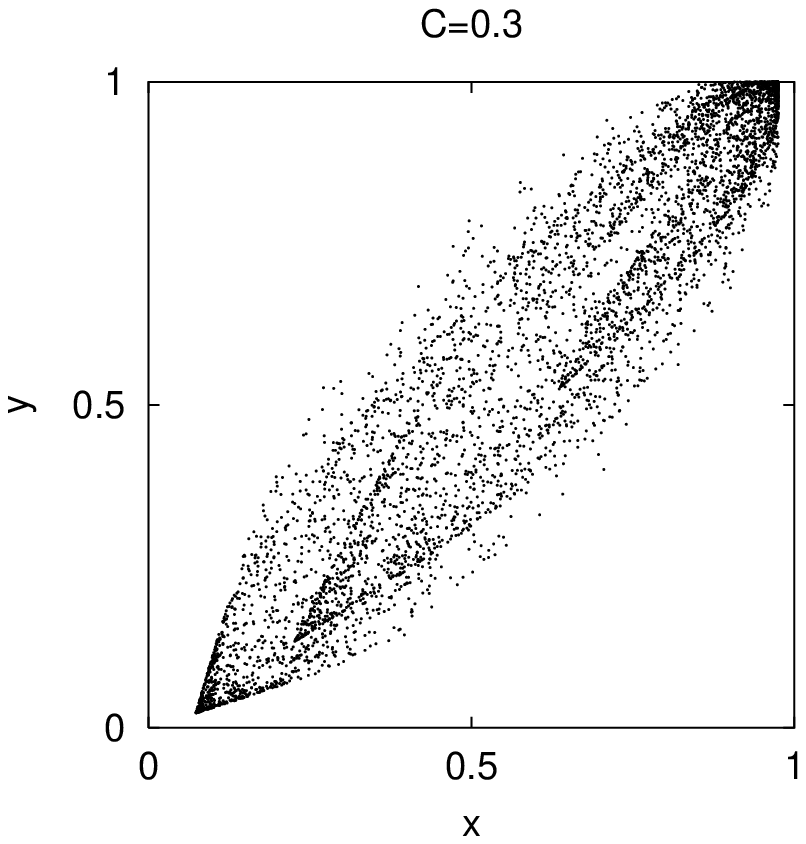}\\
    \hskip -0.5cm \includegraphics[scale=0.5]{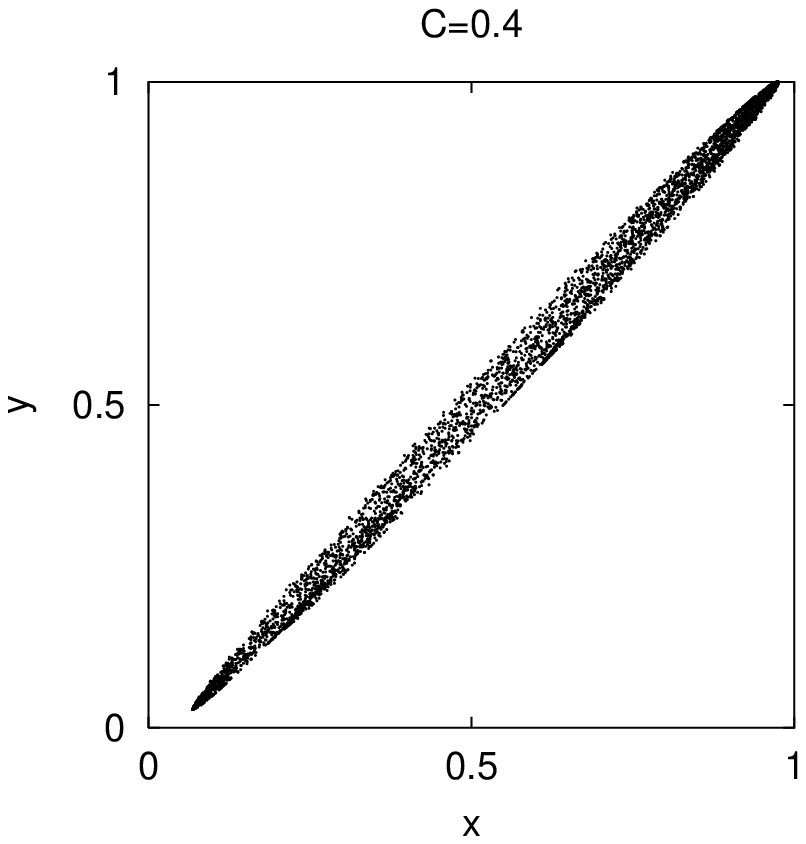}\hskip -2cm
      \includegraphics[scale=0.5]{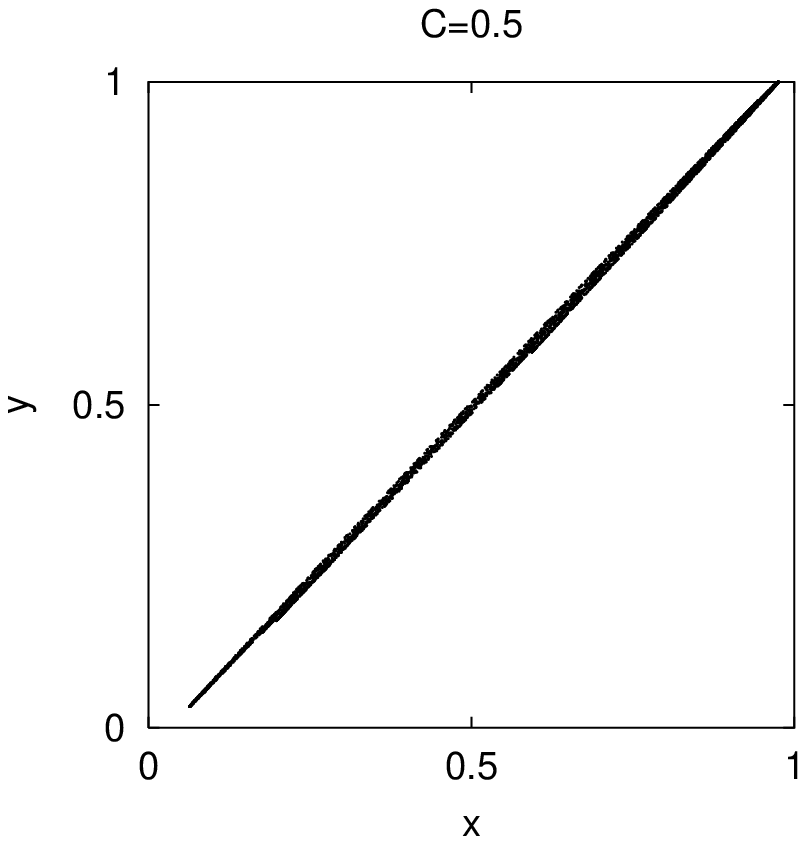}
    \end{tabular}
    \caption{Attractors of symmetrically coupled logistic maps
      (\ref{eq:couple}) for $c=0,0.1,0.2,0.3,0.4,0.5$ in $(x,y)$ plane.} 
    \label{fig:distrib}
  \end{center}
\end{figure}

 Estimated correlation dimensions for several values of the coupling
 constant $c$ are shown in figure~\ref{fig:symlogi}. One can see the
 jump in the dimension of interaction from 0 at $c=0$ to the value equal
 to the dimension of the whole systems for positive $c$ indicating
 case 4. in our classification. For $c=0.2$ asymptotic dynamics
 settles on a periodic orbit leading to all the dimensions equal to
 0. Numerically obtained approximations to asymptotic measures for
 coupling constant $c=0.,0.1,0.2,0.3,0.4,0.5$ are shown in
 figure~\ref{fig:distrib}. Note the increasing synchronization between
 $x$ and $y$. 

 It is of interest to compare the values of dimensions for $c=0$ and
 1, because in both cases $D_1(\mu_x)\approx D_1(\mu_y)=1$, but the
 dimension of the whole system, estimated from $f(x_n,y_n)$ is equal
 to 2 in the first case, and 1 in the second, implying $D_{\sf int}=0$
 and 1 in these cases, respectively. Thus the first measure has a
 product structure, while the other is concentrated on the diagonal
 $x=y$. 

The last case considered is that of the double control:
\begin{equation}
\left\{
  \begin{array}[c]{rcl}
    x_{n+1} & = & f_\alpha(x_n),\\
    y_{n+1} & = & \frac{f_\beta(y_n)+c_1 x_n}{1+c_1},\\
    z_{n+1} & = & \frac{f_\gamma(z_n)+c_2 x_n}{1+c_2},
  \end{array}
\right.
\label{eq:asym}
\end{equation}
where $\alpha=4.0$, $\beta=3.8$, $\gamma=3.9$, $c_1,c_2\in [0,1]$.
Let the observed systems $U_1$ and $U_2$ be the sets of all pairs
$(x,y)$ and $(x,z)$, respectively. Then we have essentially the case
2. If $U_1$ and $U_2$ are the sets of all points $x$ and pairs $(x,z)$
then we have the case 3.

\begin{figure}[htbp]
  \begin{center}
    \begin{tabular}[c]{c}
      (a) \hfill\ \\
      \includegraphics[scale=0.6]{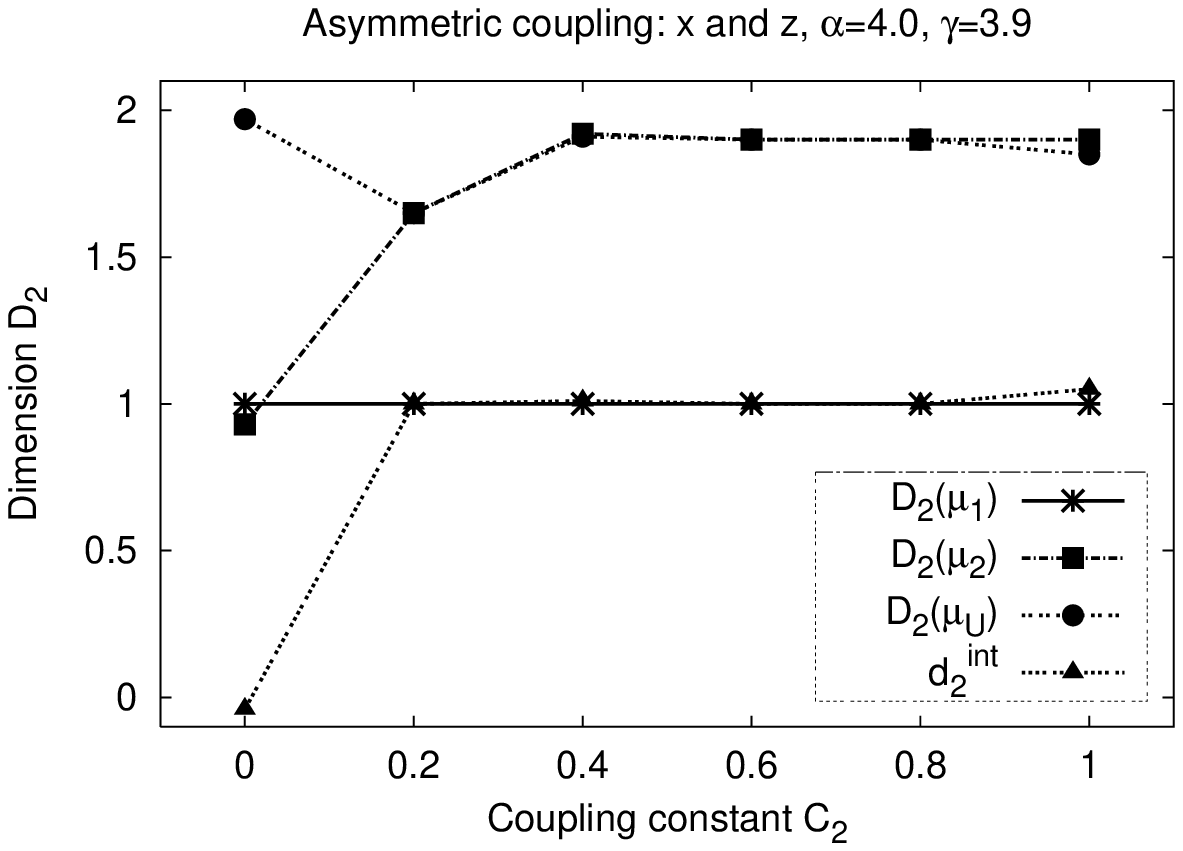}\\
      (b) \hfill\ \\
      \includegraphics[scale=0.6]{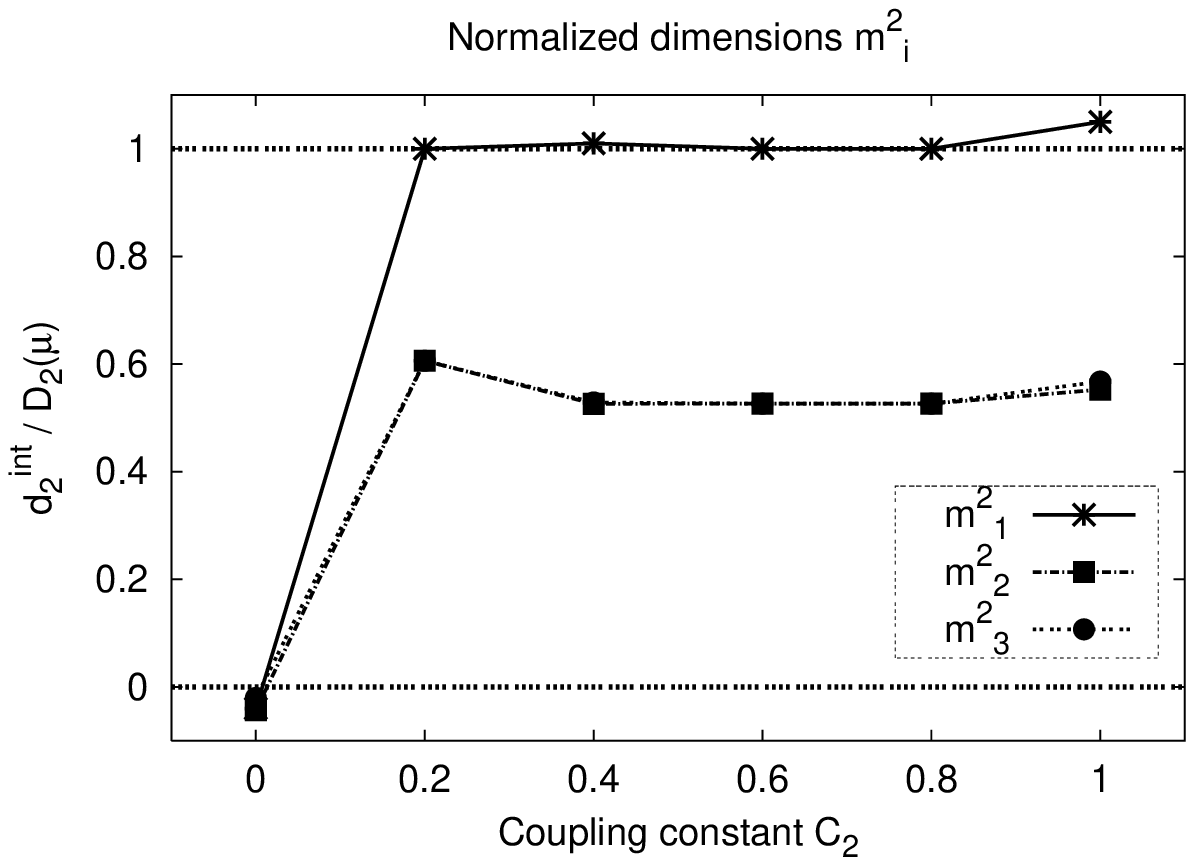}
    \end{tabular}
    \caption{a) Dimensions $D_2(\mu_1), D_2(\mu_2),D_2(\mu_U)$ and
      $d_2^{\sf int}$ of asymmetrically coupled logistic maps
      (\ref{eq:asym}) when $x$ and $z$ are the observed variables.
      b) Normalized
      dimensions $m_1^2$, $m_2^2$ and $m_U^2$ for the same systems.}
      \label{fig:asymlogi1}
  \end{center}
\end{figure}
\begin{figure}[htbp]
  \begin{center}
    \begin{tabular}[c]{c}
      (a) \hfill\ \\
      \includegraphics[scale=0.6]{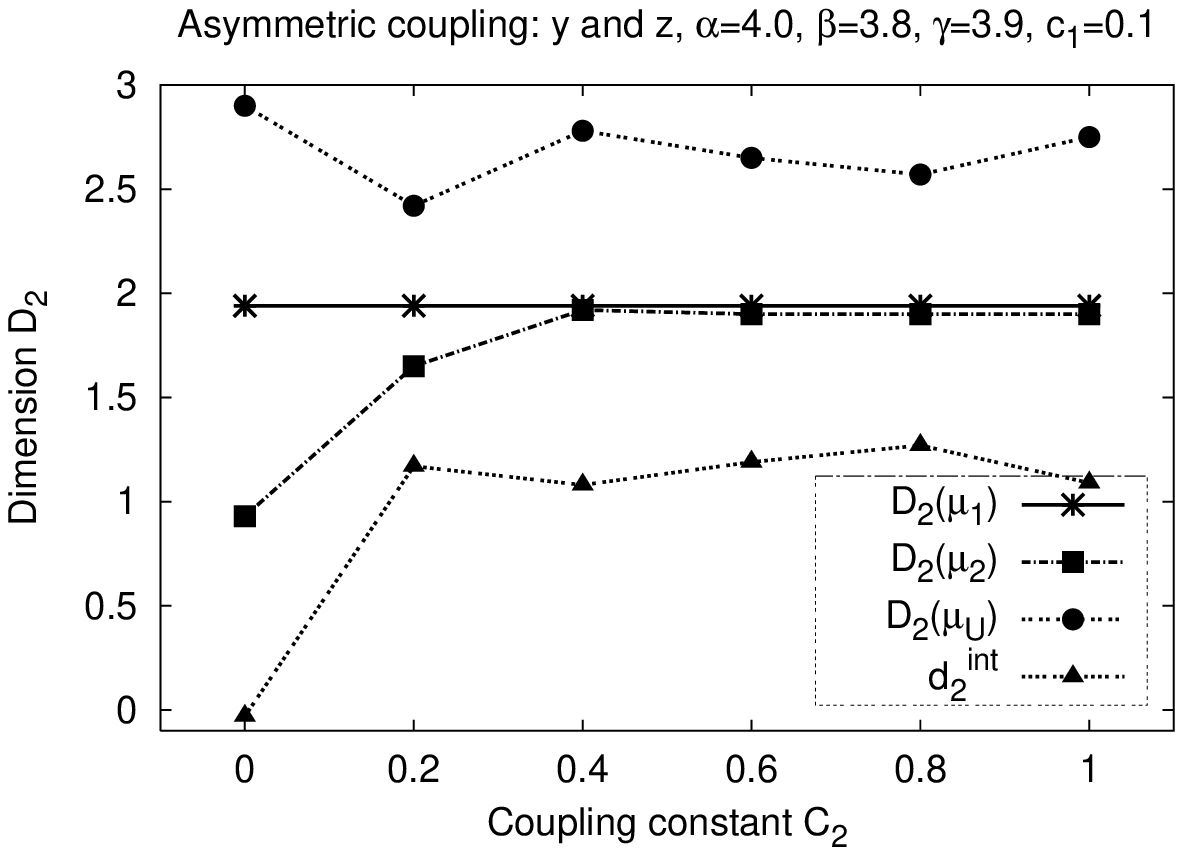}\\
      (b) \hfill\ \\
      \includegraphics[scale=0.6]{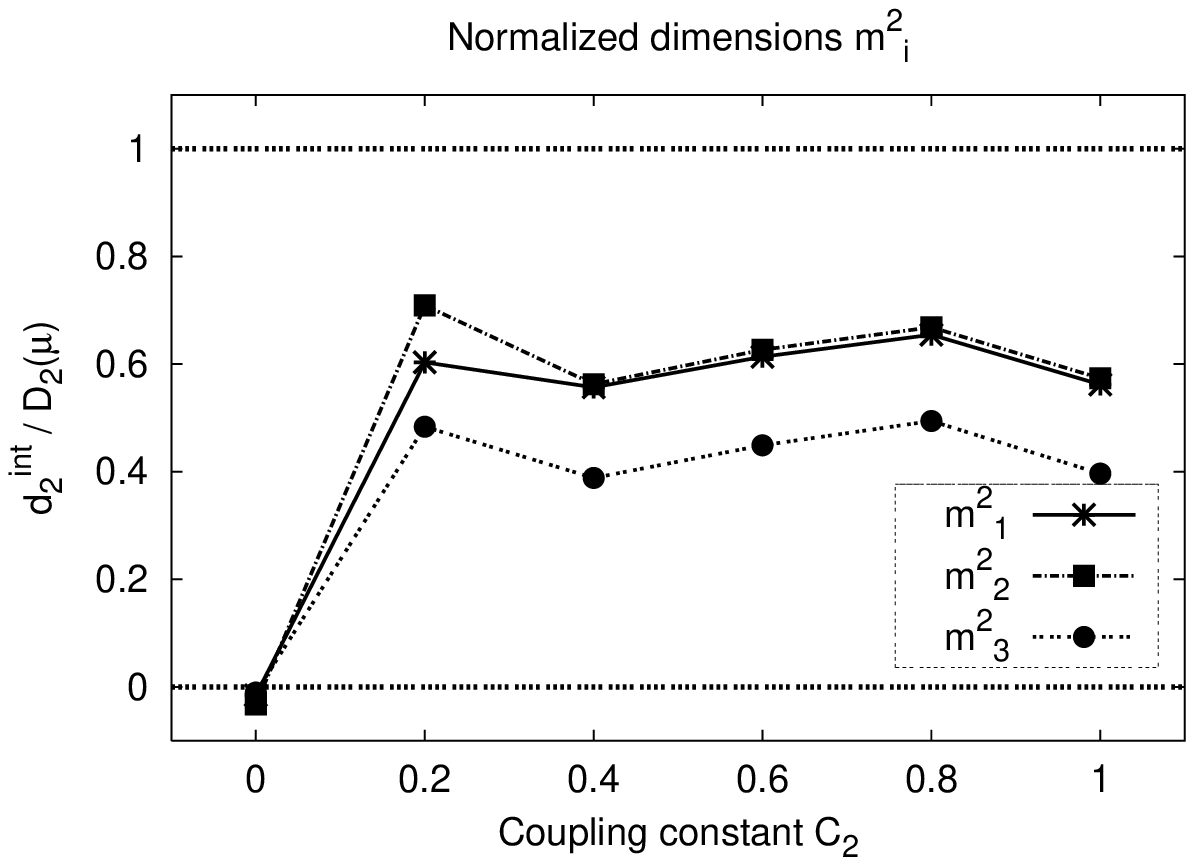}
    \end{tabular}
    \caption{a) Dimensions $D_2(\mu_1), D_2(\mu_2),D_2(\mu_U)$ and
      $d_2^{\sf int}$ of asymmetrically coupled logistic maps
      (\ref{eq:asym}) when $y$ and $z$ are the observed variables.
      b) Normalized
      dimensions $m_1^2$, $m_2^2$ and $m_U^2$ for the same systems.}
    \label{fig:asymlogi2}
  \end{center}
\end{figure} 

\begin{figure}[htbp]
  \begin{center}
    \leavevmode
    \begin{tabular}[c]{c}
      \hskip -0.5cm
      \includegraphics[scale=0.5]{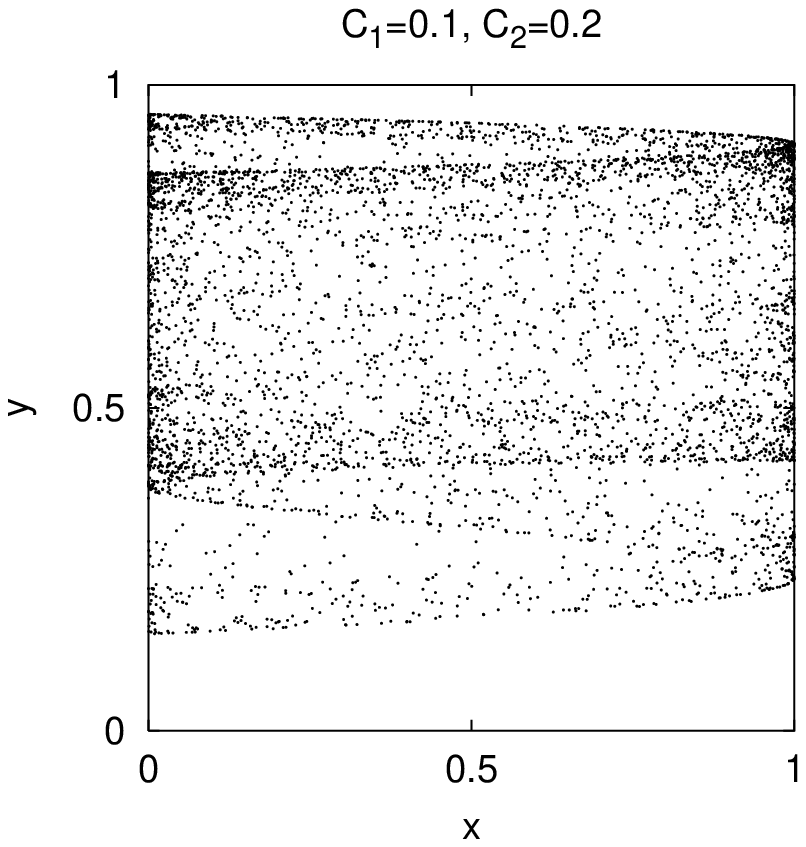}
      \hskip -2cm
      \includegraphics[scale=0.5]{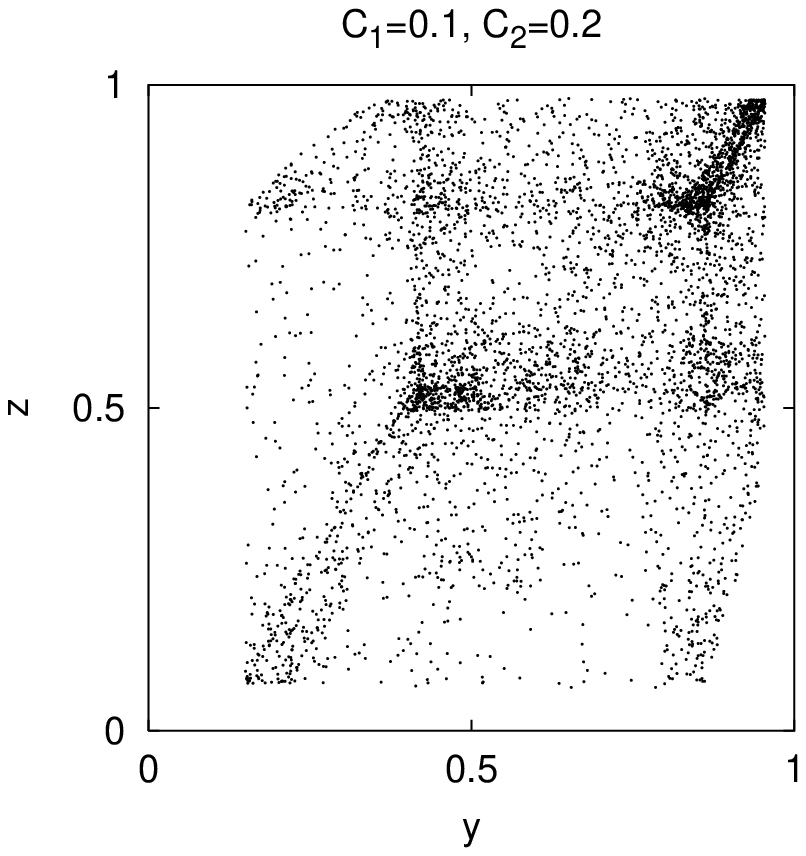}
    \end{tabular}
    \caption{Projections of the attractor of asymmetrically coupled
      logistic maps (\ref{eq:asym}) for $c_1=0.1,\,c_2=0.2$ on $(x,y)$ 
      and $(y,z)$ planes.}  
    \label{fig:doubledist}
  \end{center}
\end{figure}

Figures~\ref{fig:asymlogi1} and~\ref{fig:asymlogi2} show estimated
correlation dimension in these cases. Again, one can clearly see the
difference between the coupled ($c_i>0$) and uncoupled ($c_i=0$)
systems, because the interaction dimension jumps from 0 to 1 or more,
in agreement with our expectations from theorems 2 and 3, since the
dimension of the common part is 1 ($x_n$ evolves according to Ulam
map: $\alpha=4.0$). Figure~\ref{fig:doubledist} shows projections of
the attractor  of (\ref{eq:asym}) on $(x,z)$ and $(y,z)$ planes for
$c_1=0.1$ and $c_2=0.2$.

\section{Conclusions and outlook} 

We presented a method which allows to distinguish
interacting from non-interacting systems when time series of
variables of the two systems are available. Partial proof of its
validity was 
provided. Classification of all the possible interaction schemes was
presented with examples of all the cases. Several simple interacting
systems were analyzed.

To use our method in practice (from field data) we suggest the
following procedure: 
\begin{enumerate}
\item[A.] calculate the dimensions $D_q(\mu_1),D_q(\mu_2),D_q(\mu_U)$ and
  $d_q^{\sf int}$~(\ref{eq:dint}) (we suggest $q=1$ or $q=2$; it is
  also good to normalize the data if they are od different orders);
\item[B.] repeat the calculation for several different coupling functions
  $Y$ and average the results (linear combination seem to be the best
  choice); 
\item[C.] if they are different from 0, calculate the normalized dimensions
  $m_i^q$~(\ref{eq:normdim}). 
\item[D.] they may take one, two or three distinct values. 
  \begin{enumerate}
  \item if all of them are 0, the systems do not interact (case 1.);
  \item if all of them are greater than 0 and less than 1, this is a
    generic case of interacting systems (case 4.); 
  \item if one of them is 1, the other are smaller, all the degrees of
    freedom of one system couple to some degrees of freedom of the other 
    (case 3.), or we have the previous case (case 4.) but the
    variables of one of the systems which are not coupled to the other
    synchronize to the system comprising the common part of the
    dynamics;  
  \item if they are all equal to 1, all the degrees of freedom of one
    system couple to all the degrees of freedom of the other (case
    2.), or we have the two previous cases (3. or 4.) but the
    variables of the two systems which are not coupled synchronize to
    the system comprising the common part of the
    dynamics.  
  \end{enumerate}
\end{enumerate}

Our method has been successfully used to distinguish between
interacting and non-interacting Chua systems in an 
experiment\cite{wojcik00db}. We hope it shall prove a useful tool in
analysis of other complex systems.

\section*{Acknowledgements} Discussions with several people enriched
our understanding of the problem. In particular we want to thank Lou
Pecora, Piotr Szymczak and  Karol \.{Z}yczkowski for illuminating
comments. This work has been supported by the Polish Committee of
Scientific Research under grant nr 2 P03B 036 16.

\appendix
\section{The Proofs.}
\label{sec:appa}

Let $\mu_1$, $\mu_2$ be the invariant measures of systems $U_1$, $U_2$
as defined in Section~\ref{sec:nonsyst}.

\smallskip 

{\bf Theorem 1}
{\em Suppose $D_q(\mu_1),D_q(\mu_2),D_q(\mu_1\times \mu_2)$ exist. Then  
\[
  D_q(\mu_1\times\mu_2)=D_q(\mu_1)+D_q(\mu_2).
\]
}
{\em Proof:}
  Take $q\neq1$. For every $\varepsilon>0$ consider partitions of ${\mathbb{R}}^{n_i}$ into
  cells of volume $\varepsilon^{n_i}$. This gives a partition in
  ${\mathbb{R}}^{n_1+n_2}$ into boxes of volume $\varepsilon^{n_1+n_2}$.
  
  Let 
  \begin{eqnarray*}
    p_j & = & \mu_1(j-\mbox{\rm th cell from the cover of}\; U_1),\\
    r_k & = & \mu_2(k-\mbox{\rm th cell from the cover of}\; U_2).
  \end{eqnarray*}
  Then 
  \begin{eqnarray*}
    D_q(\mu_1\times\mu_2) & = & \lim_{\varepsilon\rightarrow0} \frac1{q-1}
    \frac{\log\sum_{k,j}p_i^q r_j^q}{\log \varepsilon} \\
    & = & \lim_{\varepsilon\rightarrow0}  \frac1{q-1}
    \frac{\log\left(\sum_k p_i^q\right) 
      \left(\sum_j r_j^q\right)}{\log \varepsilon} \\ 
    & = &  \lim_{\varepsilon\rightarrow0}  
    \left(\frac1{q-1}\frac{\log\sum_k p_k^q }{\log \varepsilon}\right) +\\
   && \lim_{\varepsilon\rightarrow0}  
    \left(\frac1{q-1}\frac{\log \sum_j r_j^q}{\log \varepsilon}\right).
  \end{eqnarray*}
  But the last two limits exist and are equal to $D_q(\mu_1)$ and
  $D_q(\mu_2)$, respectively. 

  The case of $q=1$ is straightforward and left to the reader.
{\hfill $\Box$}

\smallskip

For the next proof we need the following Lemma.

\smallskip

{\bf Lemma 5} Let $1\geq c_{ij}\geq0$, $\sum_{ij} c_{ij}=1$, $a_i=\sum_j
c_{ij}$, $b_j=\sum_i c_{ij}$. Then
\begin{equation}
  \label{inequal}
  \sum_{i,j} (a_i b_j\log (a_i b_j)-c_{ij}\log c_{ij})\leq 0.
\end{equation}

{\em Proof:}
Every convex function $f$  satisfies Jensen's inequality 
\begin{equation}
  f\left(\sum_i p_i x_i\right) \leq \sum_i p_i f(x_i),
\end{equation}
where $\sum_i p_i=1$. Since $f(x)=x \log x$ is convex, one has 
\begin{eqnarray*}
  f\left(\sum_{i,j} c_{ij}\right) &\leq& \sum_{i,j} a_i b_j f\left(\frac{c_{ij}}{a_i
    b_j}\right),\\
  f(1) & \leq & \sum_{i,j} a_i b_j \frac{c_{ij}}{a_i b_j}
  (\log c_{ij} -\log a_i - b_j)\\
  0 & \leq & \sum_{i,j}c_{ij}\log c_{ij} - \sum_{i,j}c_{ij}\log a_i - 
  \sum_{i,j} c_{ij}\log b_j\\
  0 & \leq & \sum_{i,j}c_{ij}\log c_{ij} - \sum_i a_i\log a_i -
    \sum_j b_j\log b_j\\ 
    0 & \leq & \sum_{i,j}(c_{ij}\log c_{ij} - a_i b_j \log (a_i b_j)), 
\end{eqnarray*}
where we took $p_{ij}=a_i b_j$ and $x_{ij}=c_{ij}/(a_i b_j)$.
{\hfill $\Box$}

\smallskip

Let $\mu_1$, $\mu_2$, $\mu_x$ and $\mu_S$ be the invariant measures
defined in Section~\ref{sec:intersyst}.

\smallskip

{\bf Theorem 2}
{\em Suppose $D_1(\mu_1)$, $D_1(\mu_2)$, $D_1(\mu_V)$,
  $D_1(\mu_U)$ exist. Then 
\[
    D_1(\mu_V) \leq d_{\sf int}:= D_1(\mu_1) + D_1(\mu_2) - D_1(\mu_U).
\]
(We shall call $d_{\sf int}$ {\em dimension of interaction}). The
equality holds when $y_1$ and $y_2$ are asymptotically independent. 
}

{\em Proof:}
There are $n_1+n_2$ independent variables thus the system can be
embedded in ${\mathbb{R}}^{n_1+n_2}$. Consider a partition of
${\mathbb{R}}^{n_1+n_2}$ into cells of size $\varepsilon$ consistent with
the structure of equations of dynamics, i.e. $(i,j,k)$-th cell$=A_i\times
B_j \times C_k$, where $A, B, C$ are $\varepsilon$-cells of dimension,
respectively, $k_1+k_2$, $n_1-k_1$, $n_2-k_2$ in spaces spanned by
${\mathbf x}$, ${\mathbf y}_1$ and ${\mathbf y}_2$. 

Since the dynamics of $({\mathbf x},{\mathbf y}_1)$ is independent of
${\mathbf y}_2$, the invariant measure $\mu_1(A_i\times B_j)$ can be
written as
\[
\mu_1(A_i\times B_j)=\mu_V(A_i)\mu_{(y_1|x)}(B_j|A_i)=:p_i r_{ji},
\]  
where $\mu_{(y_1|x)}(B_j|A_i)$ are the conditional probabilities of
finding the ${\mathbf y}_1$ in $B_j$ under the condition ${\mathbf x}$
being in $A_i$.   
Similarly, 
\[
\mu_2(A_i\times C_k)=\mu_V(A_i)\mu_{(y_2|x)}(C_k|A_i)=:p_i s_{ki},
\]
and
\begin{eqnarray*}
  \mu_S(A_i\times B_j \times C_k) &=&
  \mu_V(A_i)\mu_{(y_1,y_2|x)}(B_j,C_k|A_i). \\ 
 & =: &  p_i t_{jki}
\end{eqnarray*}
If $\mu_{(y_1|x)}(B_j|A_i)$ and $\mu_{(y_2|x)}(C_k|A_i)$ are independent, then 
\begin{equation}
  \label{indep}
  \mu_{(y_1,y_2|x)}(B_j,C_k|A_i)=\mu_{(y_1|x)}(B_j|A_i) \mu_{(y_2|x)}(C_k|A_i),
\end{equation}
otherwise 
the only thing we know is that the l.h.s. measure is the coupling
of the r.h.s. measures, namely
\begin{eqnarray*}
  \sum_k \mu_{(y_1,y_2|x)}(B_j,C_k|A_i) & = & \mu_{(y_1|x)}(B_j|A_i),\\
  \sum_j \mu_{(y_1,y_2|x)}(B_j,C_k|A_i) & = & \mu_{(y_2|x)}(C_k|A_i),
\end{eqnarray*}
or
\begin{eqnarray*}
  \sum_k t_{jki} & = & r_{ji},\\
  \sum_j t_{jki} & = & s_{ki}.
\end{eqnarray*}
Of course,
\[
\sum_k s_{ki} = \sum_j r_{ji} = \sum_{jk} t_{jki}=1,
\]
if $p_i\neq 0$. Otherwise we take $\forall j,k:\, t_{jki}=0$.

Taking this into consideration, inequality (\ref{eq:inter}) follows: 
\begin{eqnarray*}
  \lefteqn{D_1(\mu_1) + D_1(\mu_2) - D_1(\mu_V)-D_1(\mu_U) =} \\
  & &  \lim_{\varepsilon\rightarrow0} 
  \frac{\sum_i\sum_j
    p_i r_{ji} \log(p_i r_{ji})
    }{\log \varepsilon} +\\
  & &  \lim_{\varepsilon\rightarrow0} 
  \frac{\sum_i\sum_k
    p_i s_{ki} \log(p_i s_{ki})
    }{\log \varepsilon} +\\
  & & - \lim_{\varepsilon\rightarrow0} 
  \frac{\sum_i
    p_i  \log(p_i)
    }{\log \varepsilon}+ \\
  & & - \lim_{\varepsilon\rightarrow0} 
  \frac{\sum_{i,j,k}p_i t_{jki} 
    \log(p_i t_{jki} )}{\log \varepsilon}\\
 & = & \lim_{\varepsilon\rightarrow0} \frac{\sum_i
    p_i  \log(p_i)
    \left(
      \sum_j r_{ji} + \sum_k s_{ki} - 1 - \sum_{j,k} t_{jki}  
    \right)
    }{\log \varepsilon} + \\
  & & \lim_{\varepsilon\rightarrow0} 
  \frac{   
    \sum_i p_i
    \sum_{j,k} (r_{ji} s_{ki} \log(r_{ji}s_{ki})-t_{jki} \log( t_{jki}))
    }
  {\log \varepsilon} \\
    & \geq &  0
\end{eqnarray*}
where in the last line we used Lemma 5 for $c=t$, $a=r$ and $b=s$ and
the fact that $\log \varepsilon < 0$.

Note that the equality holds if and only if 
\begin{equation}
\label{eq:asymptindep}
t_{jki}= r_{ji} s_{ki}.
\end{equation}
This is what we call asymptotical independence of variables
$y_1$ and $y_2$. In particular, when $y_i$ are in generalized
synchrony with $x$, this means that their asymptotic behavior is
independent of their initial states and depends only on initial state
of $x$, therefore their probability distributions cannot be
independent, since they depend on the same number $x(0)$. However, we
are not sure if this the only case when the equality is not satisfied,
this is why we use another name for the above condition.
{\hfill $\Box$}

\smallskip

One would like to establish a similar inequality in case of other Renyi
dimensions, however, in general, even when~(\ref{eq:asymptindep}) is
satisfied,  
\[
D_q(\mu_S) \neq D_q(\mu_1) + D_q(\mu_2) - D_q(\mu_x).
\]
Indeed, 
\begin{eqnarray}
  \lefteqn{D_q(\mu_1) + D_q(\mu_2) - D_q(\mu_x) - D_q(\mu_S)=}
  \label{eq:dqdiff} \\
  & = & \lim_{\varepsilon\rightarrow0} 
  \frac{1}{\log\varepsilon}
  \log\left[
    \frac{
      \left(\sum_{i,k}p_i^q r_{ki}^q \right)
      \left(\sum_{l,j}p_l^q s_{jl}^q \right)
      }{
      \left(\sum_{l}p_l^q  \right)
      \left(\sum_{i,j,k}p_i^q s_{ji}^q r_{ki}^q \right)
      }
  \right] \nonumber \\
  & = & \lim_{\varepsilon\rightarrow0} 
  \frac{1}{\log\varepsilon}
  \log\left[1+
    \frac{
      \sum_{i<l,j,k}p_i^q p_l^q (r_{ki}^q-r_{kl}^q)
      (s_{ji}^q-s_{jl}^q)
      }{
      \sum_{i,j,k,l}p_i^q p_l^q s_{ji}^q r_{ki}^q 
      }
  \right].\nonumber
\end{eqnarray}
This may have arbitrary sign and needs not vanish in the limit.

Although (\ref{eq:dqdiff}) must go to 0 in the limit $q\rightarrow1$, one can
perhaps construct examples of measures for which the slope can be
arbitrarily large. On the other hand, we believe such measures will not be
typically observed in physical systems.

\section{An example of partially coupled systems.}
\label{sec:appb}

We present here a simple example of interacting systems for which one
can introduce the natural decomposition~(\ref{eq:doubledrive}).

Consider two systems $U_1$, $U_2$ interacting through a thin contact
layer $V$. Denote variables in $U_1$ as ${\mathbf u}_1=({\mathbf
  v}_1,{\mathbf w}_1)$, variables in $U_2$ as ${\mathbf u}_2=({\mathbf
  v}_2,{\mathbf w}_2)$, and variables of the contact layer $V$ are
$({\mathbf v}_1,{\mathbf v}_2)$. 
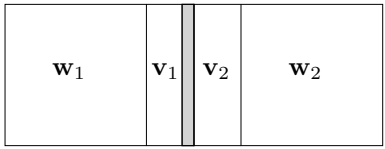
\begin{figure}[htbp]
  \begin{center}
    \leavevmode 
    \setlength{\unitlength}{0.00041666in}
    \begingroup\makeatletter\ifx\SetFigFont\undefined%
    \gdef\SetFigFont#1#2#3#4#5{%
      \reset@font\fontsize{#1}{#2pt}%
      \fontfamily{#3}\fontseries{#4}\fontshape{#5}%
      \selectfont}%
    \fi\endgroup%
    {\renewcommand{\dashlinestretch}{30}
      \begin{picture}(4824,1839)(0,-10)
        \path(312,1212)(312,1212)(312,1212)
        (312,1212)(312,1212)
        \path(12,1812)(2412,1812)(2412,12)
        (12,12)(12,1812)
        \path(3012,1812)(3012,12)
        \path(1812,1812)(1812,12)
        \path(2412,1812)(4812,1812)(4812,12)
        (2412,12)(2412,1812)
        \texture{0 115111 51000000 444444 44000000 151515 15000000 444444 
          44000000 511151 11000000 444444 44000000 151515 15000000 444444 
          44000000 115111 51000000 444444 44000000 151515 15000000 444444 
          44000000 511151 11000000 444444 44000000 151515 15000000 444444 }
        \shade\path(2262,1812)(2412,1812)(2412,12)
        (2262,12)(2262,1812)
        \path(2262,1812)(2412,1812)(2412,12)
        (2262,12)(2262,1812)
        \put(612,912){\makebox(0,0)[lb]{\smash{{{\SetFigFont{10}{24.0}{\rmdefault}{\mddefault}{\updefault}$
                  {\mathbf w}_1$}}}}}
        \put(1872,912){\makebox(0,0)[lb]{\smash{{{\SetFigFont{10}{24.0}{\rmdefault}{\mddefault}{\updefault}$
                  {\mathbf v}_1$}}}}}
        \put(3612,912){\makebox(0,0)[lb]{\smash{{{\SetFigFont{10}{24.0}{\rmdefault}{\mddefault}{\updefault}$
                  {\mathbf w}_2$}}}}}
        \put(2522,912){\makebox(0,0)[lb]{\smash{{{\SetFigFont{10}{24.0}{\rmdefault}{\mddefault}{\updefault}$
                  {\mathbf v}_2$}}}}}
      \end{picture}
      } 
    \caption{Interacting systems.}
    \label{fig:inter}
  \end{center}
\end{figure}
Dynamics of such a configuration can be described as
\begin{eqnarray*}
  \dot{{\mathbf w}}_1 & = & f_1({\mathbf v}_1,{\mathbf w}_1),\\ 
  \dot{{\mathbf w}}_2 & = & f_2({\mathbf v}_2,{\mathbf w}_2),\\ 
  \dot{{\mathbf v}}_1 & = & g_1({\mathbf v}_1,{\mathbf v}_2,{\mathbf w}_1),\\ 
  \dot{{\mathbf v}}_2 & = & g_2({\mathbf v}_1,{\mathbf v}_2,{\mathbf w}_2).
\end{eqnarray*}
If we can average the influence of ${\mathbf w}_1,{\mathbf w}_2$ on the
dynamics of ${\mathbf v}_1,{\mathbf v}_2$, e.g. when the time scales
involved in the dynamics of ${\mathbf v}_i$ and ${\mathbf w}_i$ are
different, we obtain 
\begin{eqnarray}
  \dot{{\mathbf w}}_1 & = & f_1({\mathbf v}_1,{\mathbf w}_1), \nonumber\\ 
  \dot{{\mathbf w}}_2 & = & f_2({\mathbf v}_2,{\mathbf w}_2), \nonumber\\ 
  \dot{{\mathbf v}}_1 & = & g_1({\mathbf v}_1,{\mathbf
  v}_2,\lambda_1), \label{layer} \\ 
  \dot{{\mathbf v}}_2 & = & g_2({\mathbf v}_1,{\mathbf v}_2,\lambda_2),\nonumber
\end{eqnarray}
where $\lambda_1, \lambda_2$ measure the average influence of
${\mathbf w}_1,{\mathbf w}_2$ on $V$. Thus equations for
${\mathbf v}_1,{\mathbf v}_2$ comprise a closed system $V$. This
part of dynamics is responsible for the interaction. 
Note that this scheme can also be considered as a double control
configuration of three systems, where $({\mathbf v}_1,{\mathbf v}_2)$
control ${\mathbf w}_1$ and ${\mathbf w}_2$.

If we set ${\mathbf x}:=({\mathbf v}_1,{\mathbf v}_2)$, ${\mathbf
  y}_i={\mathbf w}_i$, then the equations~(\ref{layer}) reduce to
  equations~(\ref{eq:doubledrive}).


\begin{thebibliography}{10}

\bibitem{Packard80}
N.~H. Packard, J.~P. Crutchfield, J.~D. Farmer, and R.~S. Shaw, Phys. Rev.
  Lett. {\bf 45},  712  (1980).

\bibitem{Takens80}
F. Takens,  in {\em Dynamical Systems and Turbulence (Warwick 1980)}, Vol.~898
  of {\em Lecture Notes in Mathematics ISBN 3 540 11171 9 and 0 387 11171 9},
  edited by D.~A. Rand and L.-S. Young (Springer-Verlag, Berlin, 1980), pp.\
  366--381.

\bibitem{Sauer91a}
T. Sauer, J.~A. Yorke, and M. Casdagli, J. Stat. Phys. {\bf 65},  579  (1991).

\bibitem{suave99n}
K. Suave, Consciousness and Cognition {\bf 8},  213  (1999).

\bibitem{tononi98n}
G. Tononi and M. Edelman, Science {\bf 282},  1846  (1998).

\bibitem{port95n}
R.~F. Port and T. Van~Gelder, {\em Mind as motion} (MIT Press, London, 1995).

\bibitem{thelen94n}
E. Thelen and L.~B. Smith, {\em A dynamical systems approach to the development
  of cognition and action} (MIT presss, London, 1994).

\bibitem{nowak98n}
A. Nowak and R.~R. Vallacher, {\em Dynamical social psychology} (Guilford, New
  York, 1998).

\bibitem{vallacher94n}
{\em Dynamical systems in social psychology}, edited by R.~R. Vallacher and A.
  Nowak (Academic Press, San Diego, 1994).

\bibitem{Kantz97c}
H. Kantz and T. Schreiber, {\em Nonlinear Time Series Analysis} (Cambridge
  Univ. Press, Cambridge, UK, 1997).

\bibitem{Pecora95}
L.~M. Pecora, T.~L. Carroll, and J.~F. Heagy, Phys. Rev. E {\bf 52},  3420
  (1995).

\bibitem{Schiff96}
S.~J. Schiff {\it et~al.}, Phys. Rev. E {\bf 54},  6708  (1996).

\bibitem{Rulkov95}
N.~F. Rulkov, M.~M. Sushchik, L.~S. Tsimring, and H.~D.~I. Abarbanel, Phys.
  Rev. E {\bf 51},  980  (1995).

\bibitem{Pecora97}
L.~M. Pecora, T.~L. Carroll, and J.~F. Heagy,  in {\em Nonlinear Dynamics and
  Time Series}, Vol.~11 of {\em Fields Inst. Communications}, edited by C.~D.
  Cutler and D.~T. Kaplan (American Math. Soc., Providence, Rhode Island,
  1997), pp.\ 49--62.

\bibitem{Mandelbrot82}
B. Mandelbrot, {\em The fractal geometry of nature} (Freeman, San Francisco,
  1982).

\bibitem{Meakin98da}
P. Meakin, {\em Fractals, scaling and growth far from equilibrium}, Vol.~5 of
  {\em Cambridge Nonlinear Science Series} (Cambridge University Press,
  Cambridge, 1998).

\bibitem{Pesin97dc}
Y.~B. Pesin, {\em Dimension Theory in Dynamical systems: Contemporary Views and
  Aplications}, {\em Chicago Lectures in Mathematics} (Chicago University
  Press, Chicago and London, 1997).

\bibitem{Olsen95da}
L. Olsen, Advances in Mathematics {\bf 116},  82  (1995).

\bibitem{Rnyi71}
A. R\'enyi, {\em Probability theory} (North-Holland, Amsterdam, 1971).

\bibitem{Grassberger83}
P. Grassberger, Phys. Lett. A {\bf 97},  227  (1983).

\bibitem{Hentschel83}
H.~G.~E. Hentschel and I. Procaccia, Physica D {\bf 8},  435  (1983).

\bibitem{Grassberger83c}
P. Grassberger and I. Procaccia, Physica D {\bf 9},  189  (1983).

\bibitem{Grassberger83d}
P. Grassberger and I. Procaccia, Phys. Rev. Lett. {\bf 50},  346  (1983).

\bibitem{Hausdorff19}
F. Hausdorff, Math. Annalen {\bf 79},  157  (1919).

\bibitem{Farmer83}
J.~D. Farmer, E. Ott, and J.~A. Yorke, Physica D {\bf 7},  153  (1983).

\bibitem{Farmer82c}
J.~D. Farmer, Z. Naturforsch. A {\bf 37},  1304  (1982).

\bibitem{Badii84}
R. Badii and A. Politi, Phys. Rev. Lett. {\bf 52},  1661  (1984).

\bibitem{Badii85}
R. Badii and A. Politi, J. Stat. Phys. {\bf 40},  725  (1985).

\bibitem{Grassberger85}
P. Grassberger, Phys. Lett. A {\bf 107},  101  (1985).

\bibitem{Kantz93c}
H. Kantz {\it et~al.}, Phys. Rev. E {\bf 48},  1529  (1993).

\bibitem{Halsey86}
T.~C. Halsey, M.~H. Jensen, L.~P. K.~I. Procaccia, and B.~I. Shraiman, Phys.
  Rev. A {\bf 33},  1141  (1986).

\bibitem{Falconer90}
K.~J. Falconer, {\em Fractal geometry} (Wiley, Chichester, New York, 1990).

\bibitem{Frisch85}
U. Frisch and G. Parisi,  in {\em Turbulence and predictability in geophysical
  fluid dynamics and climate dynamics}, edited by M. Ghil, R. Benzi, and G.
  Parisi (North-Holland, New York, 1985), pp.\ 84--88.

\bibitem{Paladin87}
G. Paladin and A. Vulpiani, Phys. Rep. {\bf 156},  147  (1987).

\bibitem{Tl88}
T. T\'el, Z. Naturforsch. A {\bf 43},  1154  (1988).

\bibitem{Evertsz92da}
C.~J.~G. Evertsz and B.~B. Mandelbrot,  in {\em Multifractal Measures}
  (Springer, Berlin, New York, 1992), Chap.~Appendix B, pp.\ 921--953.

\bibitem{Ott93}
E. Ott, {\em Chaos in Dynamical Systems} (University Press, Cambridge, 1993).

\bibitem{Beck97}
C. Beck and F. Schl{\"o}gl, {\em Thermodynamics of Chaotic Systems} (Cambridge
  University Press, Cambridge, UK, 1997).

\bibitem{Falconer97da}
K. Falconer, {\em Techniques in Fractal Geometry} (John Wiley and Sons, New
  York, 1997).

\bibitem{Eckmann85}
J.-P. Eckmann and D. Ruelle, Rev. Mod. Phys. {\bf 57},  617  (1985).

\bibitem{Olsen96da}
L. Olsen, Math. Proc. Camb. Phil. Soc. {\bf 120},  709  (1996).

\bibitem{Sauer93a}
T. Sauer and J.~A. Yorke, Int. J. of Bifurcation and Chaos {\bf 3},  737
  (1993).

\bibitem{Sauer97b}
T. Sauer and J. Yorke, Ergodic Th. Dyn. Syst. {\bf 17},  941  (1997).

\bibitem{Abarbanel93a}
H.~D.~I. Abarbanel, R. Brown, J.~L. Sidorowich, and L.~S. Tsimring, Rev. Mod.
  Phys. {\bf 65},  1331  (1993).

\bibitem{Abarbanel96}
H.~D.~I. Abarbanel, {\em Analysis of Observed Chaotic Data} (Springer-Verlag,
  New York Berlin Heidelberg, 1996).

\bibitem{Schreiber99}
T. Schreiber, Phys. Rep. {\bf 308},  2  (1999).

\bibitem{Casdagli91a}
M. Casdagli, S. Eubank, J.~D. Farmer, and J. Gibson, Physica D {\bf 51},  52
  (1991).

\bibitem{Wojcik00da}
D. W\'ojcik, Ph.D. thesis, Center for Theoretical Physics, Polish Academy of
  Sciences, 2000, Warsaw, Poland; in preparation.

\bibitem{hegger99da}
R. Hegger, H. Kantz, and T. Schreiber, Chaos {\bf 9},  413  (1999).

\bibitem{Takens85a}
F. Takens,  in {\em Dynamical Systems and Bifurcations, Groningen 1984},
  Vol.~1125 of {\em Lecture Notes in Mathematics}, edited by B.~L.~J. Braaksma,
  H.~W. Broer, and F. Takens (Springer-Verlag, Berlin, 1985), pp.\ 99--106.

\bibitem{Theiler88}
J. Theiler, Phys. Lett. A {\bf 133},  195  (1988).

\bibitem{Hnon76}
M. H\'enon, Commun. Math. Phys. {\bf 50},  69  (1976).

\bibitem{Pecora90}
L.~M. Pecora and T.~L. Carroll, Phys. Rev. Lett. {\bf 64},  821  (1990).

\bibitem{Pikovsky91}
A.~S. Pikovsky and P. Grassberger, J. Physics A {\bf 24},  4587  (1991).

\bibitem{Zochowski97}
M. Zochowski and L.~S. Liebovitch, Phys. Rev E {\bf 56},  3701  (1997).

\bibitem{Schuster98da}
{\em Handbook of Chaos Control}, edited by H.~G. Schuster (Wiley-VCh, Weinheim,
  New York, 1998).

\bibitem{wojcik00db}
D. W\'ojcik {\it et~al.}, in preparation (unpublished).

\end{thebibliography}
\end{document}